\documentclass[twocolumn,showpacs,preprintnumbers,amsmath,amssymb]{revtex4-1}
\usepackage{amsmath}
\usepackage{amssymb}
\usepackage{bm}
\usepackage{graphicx}
\usepackage{mathrsfs}
\usepackage{theorem}
\usepackage[normalem]{ulem}
\usepackage{xcolor}
\usepackage{natbib}

\usepackage[unicode=true,pdfusetitle,bookmarks=true,bookmarksnumbered=false, bookmarksopen=false,breaklinks=false,pdfborder={0 0 0}, backref=false,colorlinks=true,citecolor=red]{hyperref}
\usepackage{cleveref}
\newcommand{\brk}[1]{\left( #1 \right)}

\newcommand{\dif}[0]{\mathrm{d}}

\newcommand{\figref}[1]{Fig.~\ref{#1}}

\renewcommand{\eqref}[1]{Eq.~\ref{#1}}

\newcommand{\appref}[1]{Appendix~\ref{#1}}

\newcommand{\A}{\mathcal{A}}

\newcommand{\xvec}{\mathbf{x}}

\newcommand{\beq}{\begin{equation}}
\newcommand{\eeq}{\end{equation}}

\newcommand{\Kim}{K_\text{Im}}
\newcommand{\rin}{r_\text{in}}
\newcommand{\rout}{r_\text{out}}

\begin{document}
%\preprint{}

\title{Nonlinear mechanics of thin frames}


\author{Michael Moshe$^{a,b}$}
\email[]{mmoshe@syr.edu}
\author{Edward Esposito$^{c}$}
\author{Suraj Shankar$^{b,d}$}
\email[]{sushanka@syr.edu}
\author{Baris Bircan$^{e}$}
\author{Itai Cohen$^{c}$}
\email[]{itai.cohen@cornell.edu}
\author{David R. Nelson$^{a}$}
\email[]{nelson@physics.harvard.edu}
\author{Mark J. Bowick$^{b,d}$}
\email[]{bowick@kitp.ucsb.edu}

\affiliation{$^a$Department of Physics, Harvard University, Cambridge, Massachusetts 02138, USA.\\
	$^b$Physics Department and Syracuse Soft and Living Matter Program, Syracuse University, Syracuse, NY 13244, USA.\\
	$^c$Laboratory of Atomic and Solid State Physics, Cornell University, Ithaca, NY 14853, USA.\\
	$^d$Kavli Institute for Theoretical Physics, University of California, Santa Barbara, CA 93106, USA.\\
	$^e$School of Applied and Engineering Physics, Cornell University, Ithaca, NY 14853, USA.}

%\thanks{}


%\date{\today}

\begin{abstract}
	The dramatic effect kirigami, such as hole cutting, has on the elastic properties of thin sheets invites a study of the mechanics of thin elastic frames under an external load. Such frames can be thought of as modular elements needed to build any kirigami pattern. Here we develop the technique of elastic charges to address a variety of elastic problems involving thin sheets with perforations, focusing on frames with sharp corners.  We find that holes generate elastic defects (partial disclinations) which act as sources of geometric incompatibility. Numerical and analytic studies are made of three different aspects of loaded frames - the deformed configuration itself, the effective mechanical properties in the form of force-extension curves and the buckling transition triggered by defects. This allows us to understand generic kirigami mechanics in terms of a set of force-dependent elastic charges with long-range interactions.
\end{abstract}

%\pacs{61.72.-y}
%\keywords{}

\maketitle

%%%%%%%%%%%%%%%%%%%%%%%%%%%%%%%%%%%%%
% Introduction
% Putting  \emph{kirigami} in a wider context of multi-length scales problems.
\section{Introduction}
\label{sec:intro}
Classical elasticity is a scale-free continuum theory \cite{landau} and yet scale-dependent features are routinely observed in elastic materials \cite{AudolyPomeauBook}. The elastic theory of thin plates and shells \cite{Koiter1966}  is a good example: the plate/shell thickness, compared to the overall size, is a purely geometric dimensionless parameter controlling both the structural bendability and the degree of nonlinearity.  One often finds complex mechanical behavior and rich pattern formation in these structures. Thin sheets, for example, display compressional buckling \cite{landau}, wrinkling \cite{Cerda2002,Cerda2003}, and crumpling \cite{witten07}, all as a result of the interplay between the external load and the sheet thickness. 

%The classical theory of elasticity is a scale-free continuum field theory, with the appearance of scale-dependent features entirely contingent on the geometry of the elastic object and boundary conditions. A well known example is the elastic theory of thin plates and shells, where the sheet thickness provides a purely geometric length scale that dictates the global mechanics of the elastic body \cite{Koiter1966}, with plates bending easily when thin and stretching instead, when thick. The presence of a fixed \emph{intrinsic} length scale also has deep consequences for the emergence of local patterns and structures and is at the heart of many famous phenomena including the buckling, wrinkling and crumpling of thin sheets \cite{landau,Cerda1999,Cerda2003,Davidovitch13,witten07}.
% To draw a link with with non-euclidean plate theory

Non-euclidean thin sheets form another class of elastic solids characterized by multiple length scales \cite{Efrati2009}. Here the (preferred) curvature of the sheets provides an additional length scale. 
%in the problem (the thickness, the system size and the intrinsic and extrinsic curvature scales are all relevant).
Non-euclidean structures are widely prevalent in nature and play an important role in determining the morphology of flowers \cite{liang2011growth,amar2012petal}, leaves \cite{eran2004leaves,liang2009shape}, growing tissues \cite{dervaux2008morphogenesis} and seed pods \cite{Armon2011,Armon2014}. This has inspired the design of mechanically-responsive materials \cite{Kim2012,Klein2007} and actuators \cite{ionov2014hydrogel}.

\begin{figure*}
\includegraphics[width=\linewidth]{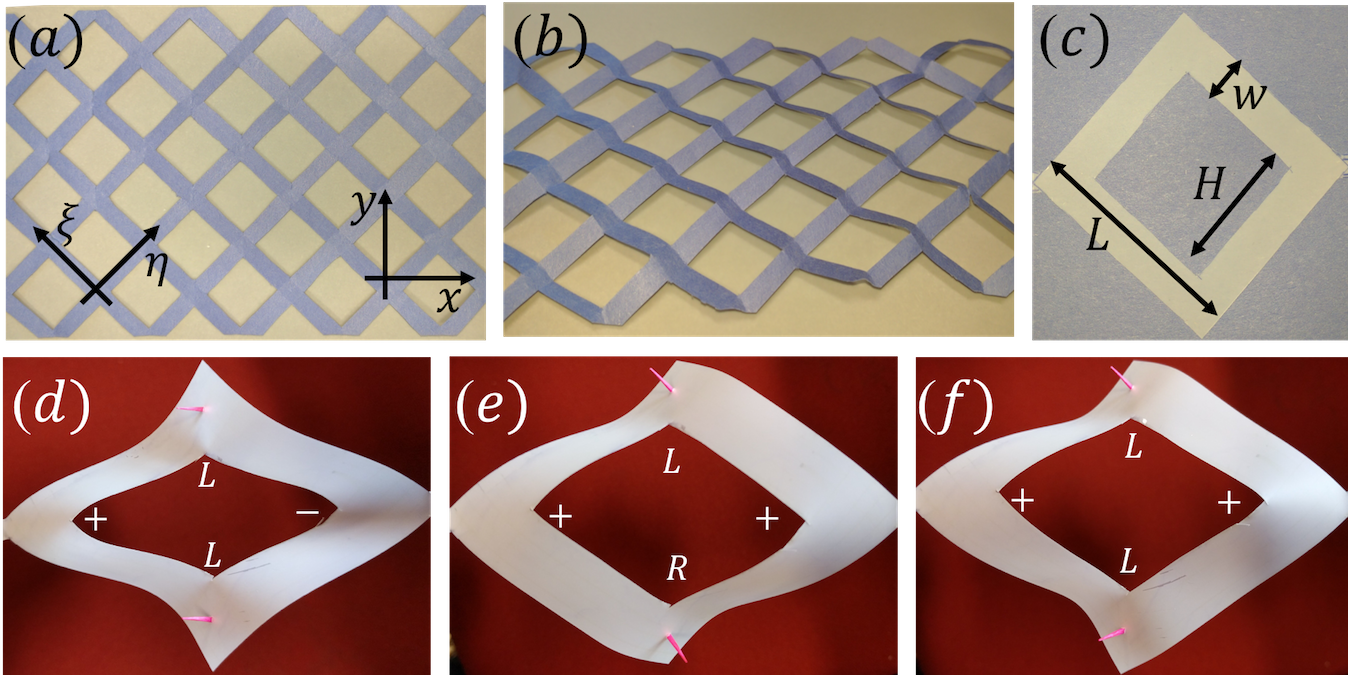}
\caption{{Kirigami and frames. (a) A periodic kirigami pattern composed of square frames. (b) The response of the pattern in (a) to a large deformation when stretched along the diagonal $x$ direction, enabled by the escape of the frame elements into the third dimension. (c) An individual square frame of size $L$, frame width $w$ and hole dimension $H=L-2w$, forming the building block of the pattern in (a). (d-f) Different locally stable configurations of a square frame subjected to an external force along its diagonal (in the horizontal direction), distinguished by the relative orientations of the four inner corners : (d) The left and right inner corners are curved up and down, respectively {(denoted by $+$ and $-$, respectively)}, whereas the top and bottom inner corners point to the left (as shown by the red needles normal to the surface {and denoted by L. This configuration is thus denoted by $+:L/L:-$}). (e) The left and right corners both point up here, while the top and bottom corners point in opposite directions, {hence denoted by $+:L/R:+$}. (f) Here both the left and right corners once again point up, but the top and bottom corners point in the same direction, {hence $+:L/L:+$}. {The remaining configuration  $+:L/R:-$ is not stable at this hole size.}}}
	\label{fig:Illustration}
\end{figure*}

% The mechanical consequences of multiscales, and demonstrate the double stranded dna
A common feature in all these examples is the presence of multiple widely-separated length scales that affect both mechanics and   structure.
%often has distinct physical consequences, showing up in observable quantities such as the total energy or the mechanical response
Such multi-scale behavior can also show up in the scaling of the energy of the system with system  size \cite{Moshe2013Nonlinearity}.
% \emph{kirigami} as another strategy for inducing a multi-scale structure, and a tool for modifying elastic properties, applicable down to the atomic scale.
From this point of view, \emph{kirigami} - the Japanese art of cutting and folding paper, is a powerful means of manipulating the geometry and the intrinsic length scales of an elastic sheet.  We find that the presence of holes provides a new handle for controlling both the onset of instabilities and the effective mechanical response. The conventional linear response of the planar state  transitions  to a mechanically softer nonlinear response as the applied force increases as a result of force-induced buckling of partial disclinations. The effective elastic properties that arise here may be tuned by varying the geometry of the holes. For large loads the displacement eventually  reaches the order of the hole size itself and we find a cross-over to a stiffer, but still nonlinear, response.  This pattern of mechanical responses, passing from a linear regime through an instability-induced softening to eventual nonlinear stiffening, is seen in other systems as well, such as
%A typical force-displacement curve for such elastic problems shows a series of transitions, starting with a regime of linear response for small displacements, followed by an instability induced softening at intermediate length scales which eventually crosses over to a stiff nonlinear response. Specific examples 
the force-induced denaturation of double-stranded DNA \cite{Smith1996}, metal alloys, solid polymeric foams \cite{Mccullough1999,Gibson1982}, and nematic elastomers \cite{Warner2003}.

%To put the focus on the  \emph{kirigami} problem:
% Literature review on recent studies related with  \emph{kirigami}
%This basic property 
The properties noted above have recently been exploited to generate mechanically actuated $3d$ configurations \cite{Dias2017,Rafsanjani2017,Zhang2015,Lamoureux2015} and highly stretchable devices \cite{Song2015,Wu2016,Shyu2015}.
\citet{Blees2015} succesfully demonstrated that kirigami can be performed at the extreme nanoscale to modify the effective mechanical properties of atomically-thin graphene in the presence of strong thermal fluctuations.  For small feature sizes, the geometry and deformation of a nanoscale graphene kirigami structure might modify its electronic transport properties as well \cite{Bahamon2016}. Lattice kirigami structures have also been used, although without direct reference to their mechanics, to create complex $3d$ macro-structures \cite{Sussman2015,Castle2016}, much in the spirit of origami-based designs. Unlike previous studies of mechanical metamaterials involving in-plane instabilities of periodically perforated thick sheets \cite{florijn2014programmable,mullin2007pattern,shim2013harnessing,matsumoto2009elastic,librandi2017porous}, we shall focus primarily on thin elastic sheets that easily buckle into the third dimension, as is most relevant for kirigami.
Given the possibilities now afforded by the design of kirigami metamaterials, an important question remains: what kinds of mechanical properties can be achieved by the techniques of kirigami? In this regard there is previous work studying the mechanical response of arrays of slits \cite{Rafsanjani2017,Shyu2015,Midori2016} and the possibility of topologically-protected floppy modes in faceted kirigami structures \cite{Chen2016}, though the full problem is far from solved.

%Decompose the problem into frames and interactions.
In search of general principles to organize the mechanics of kirigami structures, we simplify  by dividing the elastic problem into two simpler problems : the mechanics of a single frame (as illustrated in Fig.~\ref{fig:Illustration}c) which we view as a modular building block for more complex arrays  (Fig.~\ref{fig:Illustration}a), and the interaction between coupled frames. A detailed analysis of many interacting frames will be left to the future. Even the simple setting of a single frame under load is sufficient to uncover a number of general mechanical consequences of kirigami. In particular, we demonstrate that holes under load act as sources of \emph{geometric incompatibility} which in turn drive buckling. By mapping the mechanics of a pulled frame to that of a non-euclidean plate, using the formalism of elastic image charges, we are able to understand both qualitative and quantitative mechanical consequences of modifying the original geometry.  We then show through comparisons with finite-element simulations that this approach to kirigami mechanics is an improvement on simple linearized plane-elasticity methods. 

The paper is organized as follows. We start with some simple table-top demonstrations in Sec.~\ref{sec:demo} to illustrate the qualitative mechanical features of square frames. Just as electric charges tend to localize near sharp corners in conventional electrostatics, the sharp corners of square frames localize strain-dependent elastic charges in the form of buckled and unbuckled partial disclinations. In Sec.~\ref{sec:theory} we present the theoretical formalism of image elastic charges within a geometric framework of elasticity.  We then discuss the mechanics of both planar (Sec.~\ref{subsec:planar}) and buckled (Sec.~\ref{subsec:buckled}) frames. In Sec.~\ref{sec:simulation} we make a quantitative comparison of our theoretical predictions from Sec.~\ref{sec:theory} with numerical simulations and elucidate the detailed geometric dependencies in both the mechanical force response and the instability threshold.
\vspace{-1em}
\section{Table-top demonstrations}
\label{sec:demo}
Motivated by the kirigami pattern shown in \figref{fig:Illustration}(a), we choose our prototype frame geometry to be a square sheet with a centered square hole in it, whose edge length ($L$), hole size ($H$), and frame width ($w=(L-H)/2$) are shown in \figref{fig:Illustration}(c). Pulling on such a square frame along diagonally opposite ends, the first point we note is that the frame readily buckles out of the plane but can adopt multiple configurations in doing so.
%Observing the table top demonstration and extract the important phenomena for us: Localization, softening, buckling..
In Figs.~\ref{fig:Illustration}(d-f) three locally stable configurations of a diagonally loaded square frame are shown, distinguished just by the relative orientations of the buckled inner corners : Corners with angles less than $\pi/2$ in the stretched configuration (before buckling) are positive partial disclinations, and can either buckle up or down ($\pm$). Corners with angles greater than $\pi/2$ in the stretched configuration before buckling are negative partial disclinations, and the associated square plaquettes can tilt either to the left or right ($L$ or $R$). The configurations in Figs.~\ref{fig:Illustration}(d-f) can thus be compactly denoted as $+:L/L:-$ (d), $+:L/R:+$ (e), and $+:L/L:+$ (f).
%The existence of multiple (locally) energy-minimizing configurations will play an important role in understanding the effects of thermal fluctuations and entropy on \emph{kirigami} micro-structures. 

Here we primarily focus only on the global energy-minimizing configuration for a given strain and the associated energy landscape of the planar and buckled configurations for varying hole sizes and loading conditions. We neglect the effects of strong thermal fluctuations uncovered in Ref.~\cite{Blees2015}. In principle, for the square frame, there could be $2^4=16$ different buckled configurations in all, with many related by rotations and reflection symmetries. {The relative parity of opposing corners ($+$ versus $-$ and $L$ versus $R$) completely classifies the $4$ distinct buckled configurations, up to symmetry-related degeneracies.} However, the configuration $+:L/R:-$ (and its symmetry related cousins) is unstable in the parameter range we study. {The other three locally stable ones are shown in Figs.~\ref{fig:Illustration}(d-f).} We shall only briefly address some features of multi-stability in pulled frames and defer a more detailed treatment to future work. As an aside, we do note that the presence of such multiple local energy minimizers (metastable states) and their associated degeneracies would play an important role when thermal fluctuations are present, and might have nontrivial consequences for, say, the free-energy of thermalized kirigami microstructures under stress.

There are three main observations that drive our work. First, as demonstrated in \figref{fig:Illustration}(b),(d) ,the presence of a hole, or an array of holes, significantly softens  the response of a frame to external forces.  Quantifying this softening as a function of frame width, or equivalently hole size, is an important prerequisite for a thorough understanding of kirigami mechanics. Second, we find that the frame localizes curvature in the vicinity of the inner corners of a hole, much like that of a conical surface. Similar singularities and softened force-response have been observed previously in the buckling of other shapes such as slits \cite{Rafsanjani2017,Midori2016}.
Third, for small hole sizes the frame does not buckle, implying that there is a threshold hole size for buckling (at a fixed displacement). Alternately, the buckling transition may be triggered by varying the external diagonal displacement, for a given hole geometry. We shall denote the critical displacement for the buckling transition in a fixed geometry by $\delta x_c$. Guided by these observations, we now proceed to develop a theoretical framework which naturally captures and emphasizes these features of frames and kirigami.

% Theoretical model - plate theory. Equilibrium equations and assymptotic regimes (inextensible, unstretchable). Effective thickness. 
\section{Theoretical framework}
\label{sec:theory}
The mechanics of an elastic frame is governed by an elastic energy functional composed of a stretching term  depending on the the $2d$ Young's modulus $Y$ and Poisson ratio $\nu$, as well as a bending term proportional to the bending modulus $\kappa$. For a Hookean material, both the stretching and bending terms are quadratic in the stress ($\bm{\sigma}$) and extrinsic curvature (${\bf b}$) tensors, respectively. When minimizing the total energy of the system, the equilibrium equations thus obtained can be significantly simplified by using the Airy stress function $\chi$. The resulting minimization equations read in covariant form \cite{Chopin2015}
\begin{subequations}\label{eq:Equilibrium}
\begin{align}
	\frac{1}{Y}\Delta\Delta\chi&=-K\ ,\label{eq:biharmonic}\\
	\kappa\Delta\mathrm{tr}({\bf b})&=\sigma^{\mu\nu}b_{\mu\nu}\ . 
\end{align}
\end{subequations}
Here $K$ is the Gaussian curvature of the configuration adopted by the surface and is proportional to $\mathrm{det}({\bf b})$. Note that these equilibrium conditions reduce to the standard F{\"o}ppl-von K{\'a}rm{\'a}n (FvK) equations \cite{landau} upon geometrically linearizing the $3d$ configuration in a Monge patch.

The two elastic moduli together define a characteristic length scale, commonly interpreted as the effective thickness of the frame $t\equiv\sqrt{12 (1-\nu^2)\kappa/Y}$, with $\nu$ the plate Poisson ratio, typically within an order of magnitude of the actual thickness of the sheet. In the following we shall use the term ``thickness'' to mean effective thickness. The relevant dimensionless parameter that quantifies the ease with which an elastic sheet can bend rather than stretch is the F{\"o}ppl-von K{\'a}rm{\'a}n (FvK) number ($\gamma$). For a frame, as will be explained below, the appropriate definition of $\gamma$ involves the frame width $w$ as the macroscopic length scale, which gives $\gamma = Yw^2/\kappa$. When $\gamma\ll 1$, the frame typically stretches in-plane, while for $\gamma\gg 1$, it more easily trades stretching energy for bending energy and buckles out of plane instead.
%quantifying the ease with which a frame can be bent ($\gamma\ll1$) versus stretched ($\gamma\gg1$).

For small displacements, we are in the pre-buckled regime ($\delta x < \delta x_c$) and the frame remains planar (${\bf b}={\bf 0}$ and $K=0$). As discussed below, the buckling threshold $\delta x_c$ is determined by the FvK number $\gamma$. The biharmonic equation for $\chi$ along with the appropriate boundary conditions (\textit{e.g.} vanishing normal stress) completely determines the stressed state of a pulled frame. We reinterpret the solution of this problem in terms of image charges in the following section.

%The basic approach: image charges 
%Starting to simplify the problem and to extract general properties like charges, quadrupoles, etc.

\subsection{Planar frames}
\label{subsec:planar}
The primary complication in solving the plane stress problem is the presence of a non-trivial hole geometry and the corresponding boundary conditions that come with it. At this stage we note that the problem can be  solved formally using the method of image charges, often used for solving the Laplace equation in the context of classical electromagnetism \cite{jackson2007classical}. In electrostatics, the electric charge density provides a source for the Coulomb potential (via Gauss's law) which makes them dual to each other as generalized conjugate variables.
%In the same fashion,
\eqref{eq:biharmonic} tells us that the Airy stress function $\chi$ and the Gaussian curvature are related to each other in a similar fashion
%play the same role
\cite{moshe2015elastic}. This identification allows a straightforward generalization of the electrostatic image charge procedure to elastic problems. The basic idea,  a kind of variational ansatz for the frame configuration, is to guess a distribution of image ``charges'', now interpreted as sources of Gaussian curvature. This distribution determines the stress function which must also satisfy the appropriate boundary conditions on the hole. \eqref{eq:biharmonic} is then modified to be
\begin{equation}
\dfrac{1}{Y}\Delta\Delta\chi=K_\mathrm{Im}-K\ ,
\label{eq:ModifiedBiharmonic}
\end{equation}
where $K_{\mathrm{Im}}$ is the image charge induced within the hole, realized by distributing real elastic charges on its boundary. For the planar case, the Gaussian curvature vanishes ($K=0$). The distribution of image charges can then be expanded in multipoles \cite{Moshe2015PNAS}. A generic hole requires an infinite number of multipolar terms.  Topological constraints, though, require that the monopole and dipole terms in $K_\mathrm{Im}$, corresponding to a global disclination and dislocation, respectively, must vanish \cite{Kupferman2013ARMA}. The lowest order allowed multipole in $K_\mathrm{Im}$ is therefore generically the quadrupole \cite{Moshe2015PNAS}.
%\Michael{collapsed} vacancy configurations in triangular lattices can similarly be interpreted as disclination quadrupoles \cite{jain2000statistical,nelson2002defects}. 
The simple problem of a circular frame under pure external shear can thus be reinterpreted as a combination of fictitious charges at the origin and at infinity, with a solution which follows from symmetry (see \appref{app:annulus}). For a more complicated geometry and boundary conditions, such as a square hole with sharp corners pulled along the diagonal, one has to include higher order multipoles, though the quadrupole is often still the dominant contribution \footnote{For very thin frames (large holes) higher order multipole terms can be as important as the leading quadrupolar image charge. A different approach is needed in this ribbon or ring-polymer like limit (see \cite{MosheEsposito}).}. Including all the multipolar image charges is entirely equivalent to the original elastic problem and sufficient to satisfy the relevant boundary conditions. Provided the hole is not too big \cite{Note1}, this formulation characterizes perforations in an elastic sheet under stress as sources of \emph{geometric incompatibility}. 

Since \eqref{eq:ModifiedBiharmonic} is linear in $\chi$ for planar frames, we can superpose the different multipolar image charges to obtain a $K_{\mathrm{Im}}$ that satisfies the appropriate boundary conditions at the edges of the frame. The displacement field  $\vec{u}_i$ and the stress tensor  $\bm{\sigma}_i$ generated by the $i^{\mathrm{th}}$ elastic charge $\lambda_i$, is then
%We obtain expressions for the stress and displacement fields associated with the different multipoles. This allow to write a general expression for the stress function and displacement field in the form
\begin{subequations}
\begin{align}
	u^{\mu} &= \sum_{i} \lambda_iu^{\mu}_i\ ,\\
	\sigma^{\mu\nu} &= \sum_{i}\lambda_i\sigma^{\mu\nu}_i\ .
\end{align}
\label{eq:FieldsReps}
\end{subequations}
%where $\sigma_i$ and $u_i$ are the stress and displacement fields induced by the $i$'th elastic charge, and $c_i$ is its magnitude.
Note that $\vec{u}_i$ and $\bm{\sigma}_i$ are explicit functions of the individual image charges \footnote{The stress tensor $\bm{\sigma}_i$, and the corresponding strain tensor, generated by each individual charge is linear in the charge $\lambda_i$ itself for a Hookean material. On the other hand, the displacement field $\vec{u}_i$ in general depends nonlinearly on the charge $\lambda_i$ due to geometric nonlinearities in the metric tensor.  For fixed load applied at the boundary, the linear charge approximation is only valid for small charges. For for fixed external displacement, the most relevant case here, the variational problem is exactly linear in the membrane limit.} and different hole geometries only correspond to including a different number of terms and different charge magnitudes in the above sums. The functions $\vec{u}_i$ and $\bm{\sigma}_i$ can be found separately for each multipole as shown in Ref.~\cite{moshe2015elastic}.
%problem, but depends only in their specific inducing elastic charge.
Here these charges can be thought of as physically motivated variational parameters, one for each force or displacement applied along the frame diagonal.

The elastic energy in a domain $\Omega$, including forces at the boundaries, is then
\begin{equation}
E = \frac{1}{2}\int_\Omega \A_{\mu\nu\rho\sigma} \sigma^{\mu\nu}\sigma^{\rho\sigma} \mathrm{d}S - \oint_{\partial \Omega} T_{\mu} u^{\mu}  \mathrm{d}\ell 
\label{eq:FullProblem}
\end{equation}
Here $\A$ is the elastic tensor \cite{Efrati2009}, $\Omega$ is the domain of the entire frame, and $T$ is the boundary force. Writing the stress and displacement fields  in terms of the elastic charges yields 
\begin{equation}
E = \sum_{i,j} M_{ij} \lambda_i \lambda_j  - \sum_{i} m_{i} \lambda_i 
\label{eq:EnDiscrete}
\end{equation}
with
\begin{subequations}
\begin{align}
	M_{ij}&= \frac{1}{2}\int_\Omega \A_{\mu\nu\rho\sigma} \sigma_i^{\mu\nu}\sigma_j^{\rho\sigma} \mathrm{d}S\\
	m_{i}&= \oint_{\partial \Omega} T_{\mu} u_i^{\mu}  \mathrm{d}\ell
\end{align}
\end{subequations}
Since all the $\bm{\sigma}_i$ and $\vec{u}_i$ are  known explicitly, given a specific frame geometry, we can integrate over the domain $\Omega$ and obtain an expression for the matrix $M$ and the vector $m$.
%, and therefore for the elastic energy, as a function of the image elastic charges. 
After minimizing the energy with respect to the image charges $\lambda_i$, simple linear algebra leads to an explicit formula for the magnitude of the $i^{\mathrm{th}}$ charges, namely
\begin{equation}
\lambda_i = \frac{1}{2} \sum_{j}  M^{-1}_{ij} m_{j}
\label{eq:lambda}
\end{equation}
For a circular frame (i.e. an annulus) under pure shear, this method leads to the known exact result \cite{Barber1992}, with all the charges except the quadrupole and hexadecapole vanishing.
%expression for the energy as a quadratic form in the charges, and the minimization can then be done analytically.
%In \appref{} we validate this approach by comparing with the analytical solution for the circular case. 

 The more complex setup of a square frame with two localized tensions $f$ acting on diagonal corners gives rise instead to an induced fictitious quadrupole charge  
\begin{equation}
	Q(f)=\dfrac{fL}{Y}\phi_1\brk{w/L}.
	\label{eq:phi1}
\end{equation}
Here $\phi_1$ is a dimensionless rational function of the geometry, which diverges as $w\to 0$, and vanishes as $w/L \to 1/2$. The divergence as $w\to 0$ results from the softness of a very narrow (square) frame under fixed tension. For the quite different setup of a fixed displacement of the two diagonal ends of the frame, the quadrupolar charge becomes
\begin{equation}
	Q(\delta x)=L^2\phi_2\brk{w/L}\dfrac{\delta x}{L}
	\label{eq:Qdeltax}
\end{equation}
In this case the geometric function $\phi_2$ remains finite when $w\to 0$, and vanishes for $w/L\to 1/2$, as expected for an intact sheet ($w=L/2$). An explicit derivation of the geometric functions $\phi_{1,2}$ is given in the supplementary material. Similar expressions hold for higher order charges as well. Keeping the two lowest order image charges already yields an accurate approximation to the plane stress solution obtained numerically for a wide range of hole sizes (see Fig.~\ref{fig:PlanarConf} and Sec.~\ref{sec:simulation}).
%(quadrupole and hexadecapole [{\color{red} SS: Are we sure we want to use hexadecapole? In principle, two quadrupoles with the same $Q$ and alignment separated by a small distance will give this term with $4$ derivatives, so it could also be thought of as an octupole though with an unconventional construction. }])  [{\color{purple} MM: Practically the calculation assumes a hexadecapole term. Since I plan to add the mathematica file as SI, I think we must describe exactly the calculation. I guess that, at least for me, thinking of a distribution of quadrupoles may be less simple then assuming additional hexaecapolar term. }] 

Solutions parametrized with just a few image charges become more accurate for weak charges; the regime of validity of the approximation depends on the specific protocol of the prescribed deformation. For a given narrow frame ($w/L\ll 1/2$), the corresponding charge induced by a prescribed force will be larger than that induced by a prescribed displacement. A small prescribed displacement, for example, results in charges that decrease as the frame narrows. We expect (see \appref{app:annulus}) that for a fixed prescribed force, the pure quadrupole approximation is valid for $0.125\lesssim w/L<0.5$ (meaning the energy deviates from the exact value by less than $5\%$). Although we find the quadrupole approximation quite helpful for physical intuition, it will not in general satisfy the exact boundary conditions; for that, one would need to account for all possible multipoles or resort (as we do later in Sec.~\ref{sec:simulation}) to numerical calculations.

\begin{figure}
	\includegraphics[width=\columnwidth]{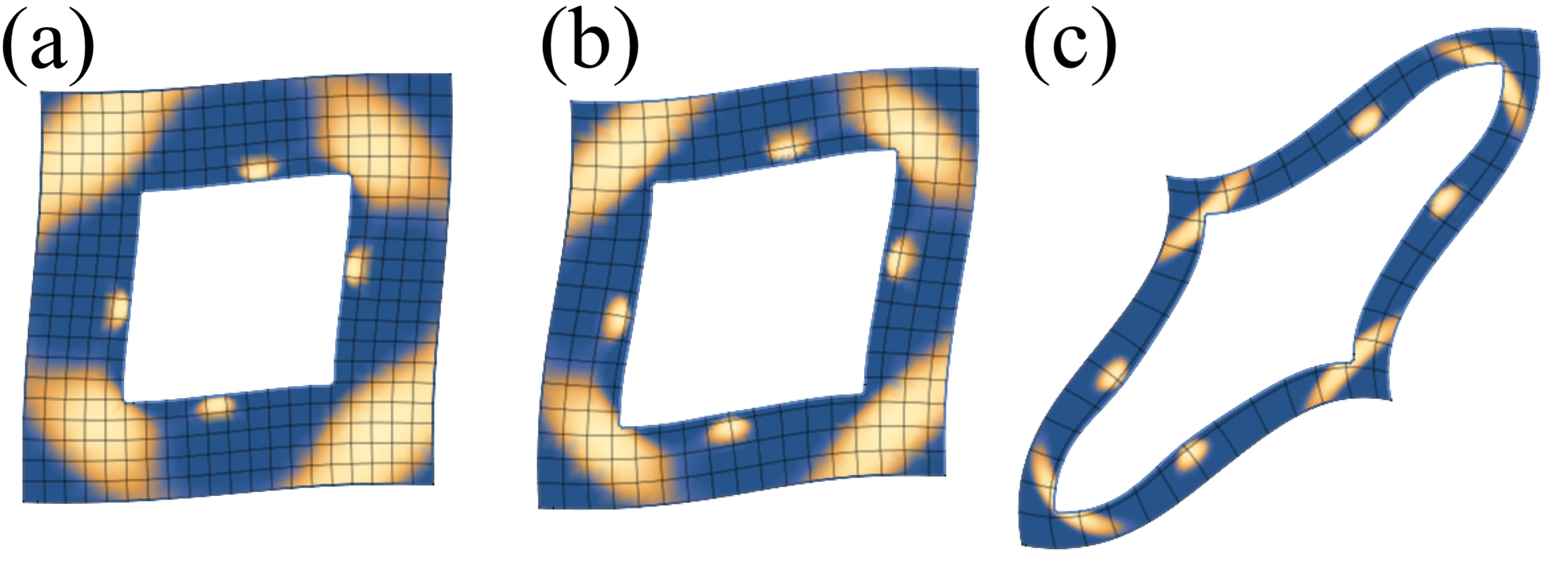}
	\caption{The deformation of a square frame constrained to be planar and subjected to a fixed force $f/(Y L)=0.1 $ along the diagonal (from the lower left to the upper right) with fixed $L=1$ but varying hole size $H=\{0.5,0.6,0.8\}$, as calculated by \eqref{eq:lambda}. The configurations are colored by the energy density on a linear scale from blue (low) to white (high). Although the force is fixed, for an appropriate choice of the effective thickness, and if allowed to escape into the third dimension, configuration (a) remains planar, (b) is planar at the buckling threshold, and (c) is beyond the critical force threshold for buckling. Note that the interior angles of the frames deviate from $\pi/2$ with the application of the force.}
	\label{fig:AnalyticConfig}
\end{figure}

Our approach also allows us to compute the deformed shape of a frame under load. By substituting the image charges from \eqref{eq:lambda} into \eqref{eq:FieldsReps} we recover the displacement field. In \figref{fig:AnalyticConfig} we use these displacements to plot configurations of planar square frames of different frame widths subjected to the same fixed force applied along the diagonal. The color encodes the stretching energy density. Although the force is constant, the resulting displacement varies dramatically as we vary the frame width: $w/L=0.25$ (a), $0.2$ (b) and $0.1$ (c) (see \figref{fig:AnalyticConfig}).

%Our original results: recovering deformation. Explaining the curvature localization.
%A main advantage of describing the solution in terms of image charges is the ability to recover the deformation field, as demonstrated in \cite{} (SI sec. ).
%Adopting this approach we calculate the deformation field induced by the image charges, and later compare with numeric solutions.  For demonstration we plot three deformed frames subjected to the same force, but have different hole sizes. The frames are colored by their energy density each with an arbitrary color-bar. 

The analogy with electrostatics provides simple interpretations of various features of these image elastic charge as well.
The hole plays a role of a conductor in electrostatics. Just as induced electric charges concentrate at regions of high curvature on conductors  in an external electric field, the elastic charges induced within a hole in response to an external load tend to localize at the sharp corners of the hole. In the planar case this is evident in energy density plots of \figref{fig:AnalyticConfig}.
In the $3d$ case, as we shall see in the next section, this elastic charge localization leads to curvature localization upon buckling, explaining the observations from the table-top experiments.

Before moving to the buckled case let's estimate the mechanical response of the frame. The geometry of the square hole allows the induced quadrupolar charge to fractionalize into four partial disclinations that localize at each corner, alternating in sign. Given this localization, the relevant length scale governing the energy of a single partial disclination is the corner plaquette size $\sim w$. We therefore estimate the total energy as that of four (unbuckled) corner disclinations:
\begin{equation}
	E_{2d}\brk{\delta x}\approx 4\frac{Ys^2}{32\pi}w^2-f\delta x\ ,
\label{eq:PlanarDisclinations}
\end{equation}
where (using Eq.~\ref{eq:Qdeltax})
\begin{equation}
	s = Q(\delta x)/H^2 = \phi_2\brk{w/L} \frac{L^2}{H^2} \frac{\delta x}{L} \equiv {\Phi}(w/L)\frac{\delta x}{L}
    \label{eq:InducedDisclination}
\end{equation}
is the partial disclination charge. Minimizing the energy with respect to $\delta x$, the effective spring constant $f=k_{\mathrm{eff}}\delta x$ of a frame that remains planar is
\begin{equation}
	k_{\mathrm{eff}}=\mathrm{const.}\ Y\left(\dfrac{w}{L}\right)^2{\Phi}(w/L)^2\ .
 \label{eq:KeffPlanar}
\end{equation}
We expect that the function $\Phi(w/L)$ is of order unity in the range $w/L \gtrsim 1/8$ where the quadrupole approximation is valid.

In summary, we have shown in this section that the solution to the elastic problem of a planar frame can be approximated using image elastic charges formed in the interior of the hole to approximately enforce boundary conditions. These image charges are a mathematical representation of the real elastic charges distributed along the boundary of the perforation, as discussed in \appref{app:annulus}. Depending on the hole shape, the induced elastic charges migrate toward highly curved regions, as suggested by the analogy with electrostatics.

\subsection{Buckled frames}
\label{subsec:buckled}
An elastic sheet embedded in a medium may also escape into the third dimension. Under load it will first stretch and then buckle beyond a threshold displacement $\delta x_c$. Let's take the case when bending is energetically much easier than stretching for a given area sheet, ($\gamma\gg 1$). In this regime the buckled configuration will be approximately stretchless. Generic configurations will be intrinsically curved  with spatially inhomogeneous Gaussian curvature, particularly on the boundaries and near the corners. Thus $K\neq 0$ in \eqref{eq:ModifiedBiharmonic} - see \appref{app:curvarture} for an illustration of distributed Gaussian curvature over a boundary under load. The stress free boundary conditions on the hole can once again be satisfied by a collection of image charges, resulting in a non-vanishing $\Kim$ as well. Because of \eqref{eq:ModifiedBiharmonic}, we expect the geometric and elastic charge distributions along the boundaries to approximately cancel to minimize energy. This requires $K=\Kim$: the Gaussian curvature of the buckled sheet screens out the sources of stress generated by the elastic charges induced on the boundaries. These elastic charges are essentially the same as those for the planar problem treated above. This screening effect resembles that found for the buckling of topological defects in crystalline membranes (see Ref.~\cite{Seung88}).

% The image charge approach affords an immediate simplification wherein the complex mechanical properties of the buckled configuration can be directly inferred by virtue of screening the induced image charge, which itself is computed in the simpler planar problem.
% Unlike the planar case, here the screening strongly depends on the precise geometry, especially the hole size.

Since screening charges localize near the inner corners of the frame, we expect that %In \figref{fig:Figure1}(e) the blue arcs mark
each fractionalized partial disclination deforms the sheet on a scale of size $w$ (just as in the planar case). If $w/L\lesssim 1/4$ the partial disclinations are separated by distances larger than $w$ and we expect their interactions to be negligible. In this case, we can estimate the energy of the $3d$ buckled configuration simply as a superposition of conical energies at each corner.
%the formation of the appropriate screening curvature.
In addition, if $w/L\gtrsim 1/8$, we find (see below) that the solution is well-approximated by a single quadrupole and we can neglect higher multipole contributions. This allows us to distinguish three main geometric regimes: (i) $w/L>1/4$ - small hole sizes where the interaction between fictitious partial disclinations is important; (ii) $1/8<w/L<1/4$ - intermediate hole sizes where the deformation is well-described by a single fictitious quadrupole and noninteracting partial disclinations and (iii) $w/L<1/8$ - large hole sizes where higher order multipoles become important.

Small holes have a minor effect on the elastic softening of a single frame. Very large holes become very flexible to bend both because they have less material  and because of their shape. In the large hole regime, the frame is more akin to a thin elastic ribbon joined at its ends (see  Ref.~\cite{MosheEsposito}). The intermediate hole size regime, probably the most accessible for exploring kirigami mechanics, will be our main focus below. 

The energy of the buckled $3d$ configuration when subjected to an external force $f$ is estimated following arguments similar to those used in the planar case. Unlike the planar case, however, the buckled quadrupoles controlling force-extension curves have multiple locally stable configurations corresponding to various combinations of up-down buckling of each partial disclination (see \figref{fig:Illustration}).  Each of these states may have a different energy. The energetic difference between the solutions originates from the interactions between the buckled partial disclinations, which we neglect to a first approximation in the intermediate hole regime where the partial disclinations are well separated. Therefore, as in the planar case, we estimate the total energy as the sum of the conical buckling cost localized at each corner \cite{Seung88,muller2008conical,efrati2015confined}
%   With the energy localized near the corners, we find
\begin{equation}
\begin{split}
E_B\brk{\delta x} = E_{2d} \brk{\delta x_c} + 4 \kappa\ln \brk{w/r_c} \brk{c_1 \, s + c_2 \, s^2 + ...} \\ - f (\delta x - \delta x_c)\ ,
\end{split}
 \label{eq:BuckledDisclinations}
 \end{equation}
 where $r_c$ is a microscopic core size regularizing the conical singularity, and is expected to scale linearly with the effective frame thickness $t$. Here the first term is the stretching work done up until buckling.  The numerical constants $c_1,\ c_2$ depend on the precise mode by which each partial disclination buckles.  The partial disclination charge $s$ is given as before by \eqref{eq:InducedDisclination}.
Note the logarithmic dependence of the energy on the frame width $w$, which is typical for conical surfaces, and is a direct consequence of the curvature localization. We emphasize that the above energy estimate is valid only in the post-buckled regime ($\delta x\gg\delta x_c$).

Upon minimizing Eq.~\ref{eq:BuckledDisclinations} with respect to $\delta x$, we find the force-displacement relation $f\brk{\delta x}$. This gives an  effective spring constant $k_{\mathrm{eff}}=\dif f/\dif\delta x$ of the form
\begin{equation}
% F &\propto \kappa_B \ln\brk{w/r_c} \brk{c_1 + 2c_1 s + ... }\ ,\\ \Rightarrow
	 k_{\mathrm{eff}}=\mathrm{const.}\ \dfrac{\kappa}{L^2} {\Phi}\brk{w/L}^2 \ln\brk{w/r_c}
\label{eq:keffBuckledSmall}
\end{equation}

We can estimate the scaling relations associated with the buckling transition of a frame under load more precisely. The buckling transition can be crossed either by varying the hole size for a fixed external force or by increasing the force at fixed frame width. Since the buckling of a frame is controlled by the buckling of partial disclinations, we can take over results from the buckling of topological disclinations \cite{Seung88}. In Ref.~\cite{Seung88} it was shown that a single disclination of charge $s$ in a finite crystalline membrane (of linear size $\sim R$) buckles only when the defect charge exceeds a critical threshold $|s_c|=\gamma_c/\gamma$ with $\gamma_c\approx120$. Note that $\gamma=YR^2/\kappa\to\infty$ as $R\to\infty$ which means the threshold vanishes and  \emph{all} topological disclinations buckle in the thermodynamic limit.
%dimensionless F{\"o}ppl-von K{\'a}rm{\'a}n number $\gamma = s R^2 Y_{2D}/\kappa_B$ exceeded a critical threshold $\gamma_c \approx 120$. 
In  our case, this analysis yields a partial disclination charge $s=\Phi(w/L)\delta x/L$ which clearly depends on the prescribed displacement as well as the frame's geometry.
%\begin{equation}
%\gamma_\text{frame} = \Phi\brk{w/L}\frac{\delta x}{L} \frac{w^2 Y_{2D}} {\kappa_B}. 
%\end{equation}
% Repeating the same steps for our partial disclinations we find
% \begin{equation}
% 	|s_c|\sim \kappa_B/ Y_{2D}w^2 ,
% \end{equation}
% and we now denote $\gamma=Y_{2D}w^2/\kappa_B$ as the FvK number appropriate for frames.
This geometric dependence of the charge on $w/L$ has no analogue in the case studied in Ref.~\cite{Seung88}. After re-expressing the critical displacement in terms of the critical charge, we obtain
\begin{equation}
	\dfrac{\delta x_c}{L} =  \dfrac{\gamma^{\mathrm{frame}}_c\kappa_B }{Y w^2 \Phi(w/L)} = \dfrac{\gamma^{\mathrm{frame}}_c  t^2 }{12 w^2 \Phi(w/L)}  .
    \label{eq:CritDisp}
\end{equation}
where by $\gamma_c^{\mathrm{frame}}$ we denote the numerical value $\gamma_c$ corresponding to the buckling of frames. 
% Writing $F =k_{\mathrm{eff}}\delta x$, and using Eq.~\ref{eq:KeffPlanar} we can rephrase this as a critical buckling load as well
% \begin{equation}
% 	F_c \propto \log(w/r_c) \frac{\kappa_B}{L}\ .
% \label{eq:CritMed}
% \end{equation}
%For larger holes ($W/L>3/4$), the frame always ``buckles'' by twisting the side ribbons. 
%To calculate the critical force in the large hole regime we estimate the crossover energy between stretching and bending of 
%four coupled ribbons, 
%that is equating the minimizing
%\ss{we equate the minimizer of} Eq.~\ref{eq:RibbonEnergy} to Eq.~\ref{eq:PlanarDisclinations}. Using the relation between $s$ and $F$ in the planar case we find
%\begin{equation}
%	F_c \propto \frac{\kappa_B}{L}\left(\frac{w}{L}\right)
%\label{eq:CritLarge}
%\end{equation}
% Unlike the previous case of intermediate size holes, where the buckling of partial disclinations is a transition via loss of stability of the planar state, for larger holes, as the energy $E\sim\delta x^2$ in both the planar and the buckled states, we don't have a strict buckling transition.
%Another prominent difference between \eqref{eq:CritLarge} and \eqref{eq:CritMed} is the dependence \ss{of the threshold load on} the hole size. This difference reflects a deep property of the screening charges: in the intermediate hole regime the magnitude of the screening disclination is insensitive to the hole size. This is in contrast to the large hole case where the buckling mechanism is fundamentally different. 

To summarize the predictions of our theoretical analysis, we find that within the intermediate hole size regime the deformation of a frame can be accurately described using a fictitious quadrupole, whose spatial distribution depends on the precise shape and geometry of the hole. For a square hole the screening defects are localized near the sharp corners, a phenomenon that is well known in electrostatics and is consistent with the table-top demonstration presented in the Sec.~\ref{sec:demo}. This allows an estimation of the effective spring constants as well as a quantitative description of the geometry-dependent scaling of the buckling transition.
%, finding a significant difference from that of topological disclinations. 
%For larger holes, the approach of fictitious charges becomes complicated, and it is instead simpler to model the frame as coupled thin ribbons.  
%\begin{itemize}
%	\item large displacement, buckled configuration. 
%	\item Solution as screening of fictitiuos charges.
%	\item localization of charges
%	\item three regimes by charges
%	\item Estimation of the energy for the different regimes.
%\end{itemize}
%\subsection{results}
%\begin{itemize}
%	\item Numeric results for planar configuration and comparison with the theoretical calculation.
%	\item Numeric results for buckled configuration, curvatre localization.
%	\item buckling transition and dependence of critical vkn in hole size.
%	\item transition between global energy minimizer.
%\end{itemize}
%\subsection{Summary}
%\begin{itemize}
%	\item Quadrupolarisability
%\end{itemize}
\section{Numerical Simulations}
\label{sec:simulation}
\begin{figure}
	\includegraphics[width=0.9\columnwidth]{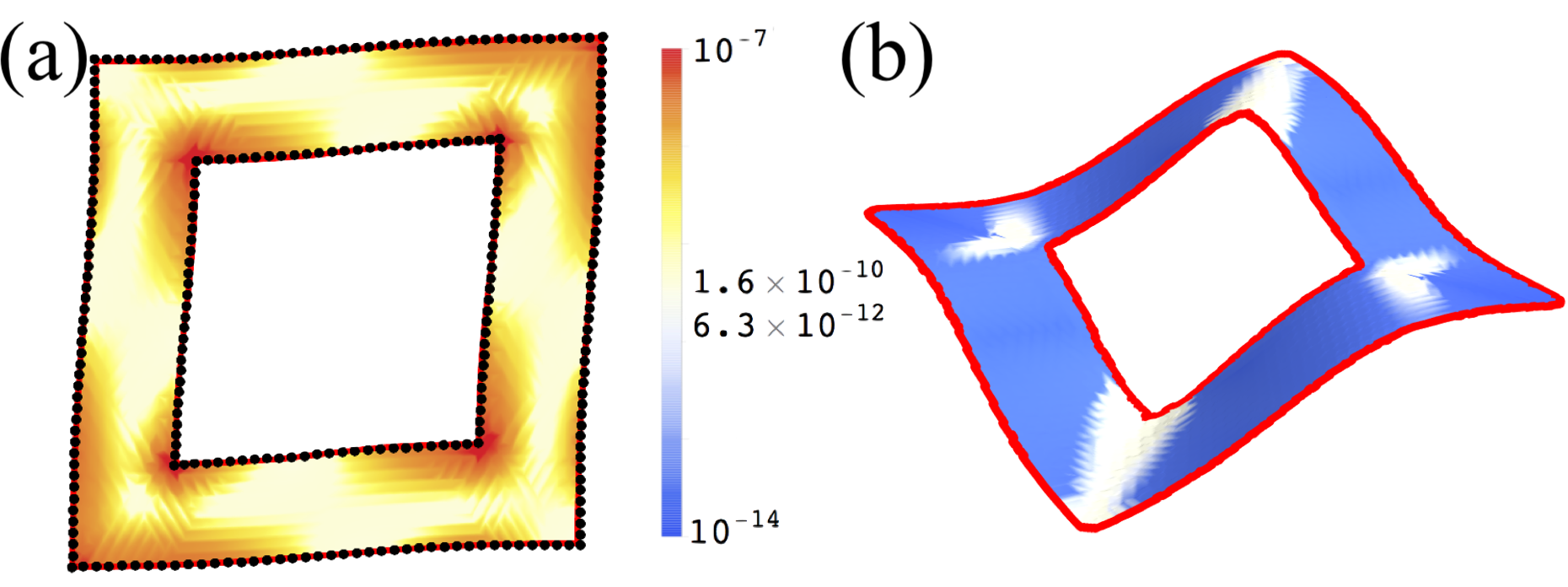}
	\caption{(a) Planar and (b) buckled, energy minimizing configurations obtained by numerical minimization of a frame of side length $L = 1$, width $w = 0.45$ and prescribed displacement $\delta x /L = 0.05 $. The frames are colored by their energy density. The scaling of the color-bar is chosen to highlight  the prominent features of the energy distribution in both configurations. The black dotted lines in (a) are the frame boundaries in the quadrupole-hexadecapole approximation described in the text. Note the close agreement between these dotted boundaries and the actual planar configuration computed numerically.}
	\label{fig:PlanarConf}
\end{figure}

To test the quadrupole and hexadecapole approximations for frames of intermediate size, we now analyze the problem numerically.
We perform finite element simulations to numerically minimize a discretized version of the elastic energy functional (as used for example in \cite{Armon2014,Therien2015}). A related method was used in Ref.~\cite{Seung88}. For each frame width a triangulation of the square frame is generated, and the energy is numerically minimized for different material parameters and external displacements.
% \begin{figure}
% 	\centering
% 	\includegraphics[width=0.9\columnwidth]{PlanarKeff.png}
% 	\caption{Comparison between theoretical and numerical effective spring constant as function of $H/L$ for planar frames- no fitting parameters}
%     \label{fig:PlanarKeff}
% \end{figure}

In \figref{fig:PlanarConf}(a) we show a representative equilibrium configuration of a planar frame under a fixed diagonal displacement. The frame is colored by its elastic energy density, confirming our expectation of energy localization near the inner corners. The black dotted lines plotted on top of the configuration's boundaries are the analytical calculations for the deformed frame, using the the two lowest order multipoles, with the quadrupole and hexadecapole charges as variational parameters. Note the good agreement. Although we have not succeeded in producing the analogue of the quadrupolar approximation for the buckled states, the corresponding disclination charges nevertheless provide insight into the buckled state.
\begin{figure}
	\includegraphics[width=0.75\linewidth]{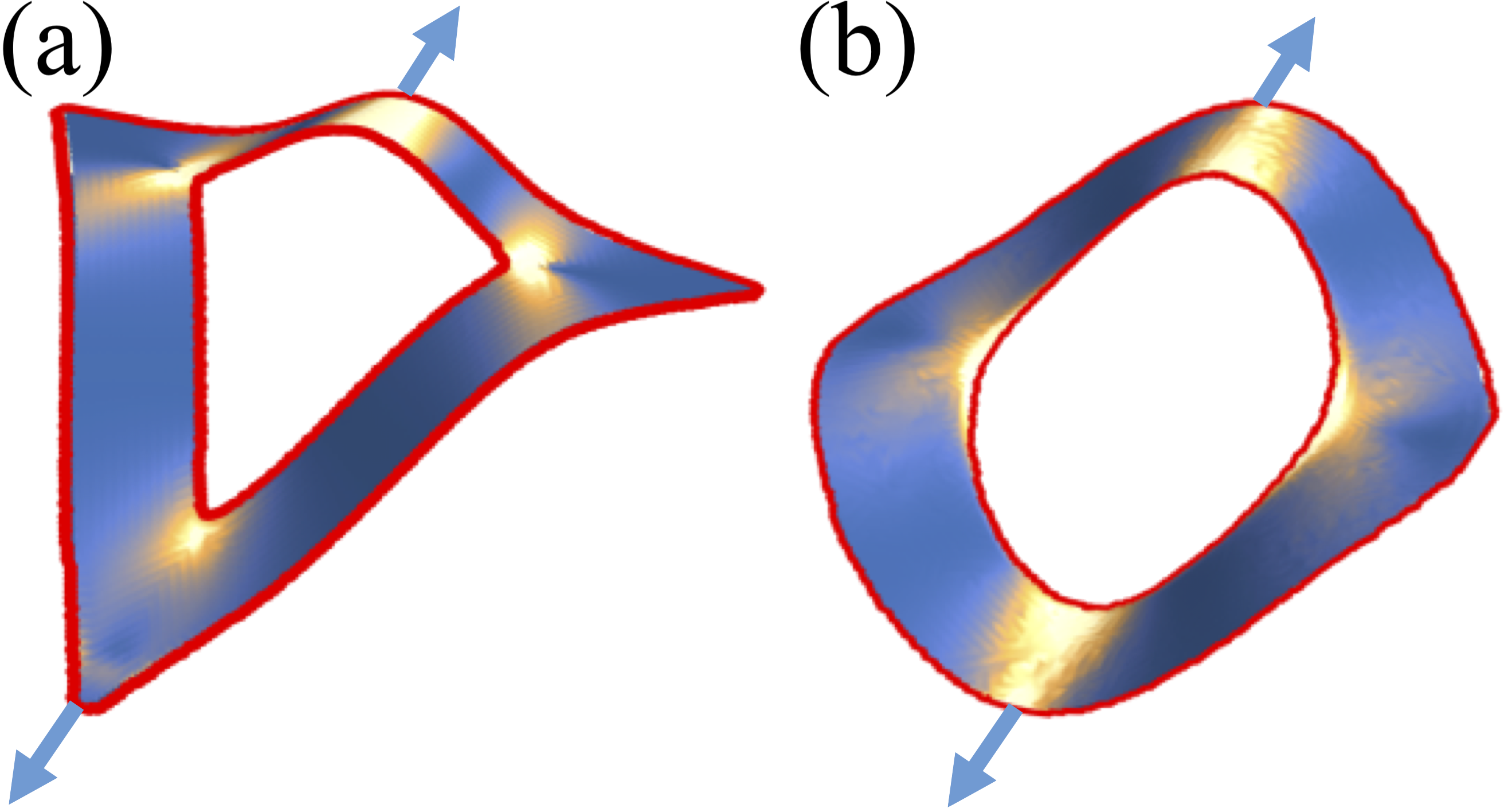}
	\caption{Three dimensional energy minimizing configurations after buckling, obtained by a numerical analysis of the full elastic problem. (a) Triangular and (b) circular frames subjected to uniaxial forces applied at opposite ends. The screening charges responsible for these energy distributions are still of a quadrupolar nature, with the partial disclinations attracted to the three sharp corners in the trianglular case, and more smeared out in the annular geometry and along one side in the case of the triangle. }
	\label{fig:Triangle}
\end{figure}

In \figref{fig:PlanarConf} (b) we present the configuration of the same frame as before, only now after allowing buckling. In this case we confirm that the stretching energy content in the system is indeed negligible compared to the bending energy after buckling. Our observations from the table-top demonstration are also confirmed - we find clear evidence of energy localization reflecting the localization of curvature near the inner corners. This feature appears naturally in our analysis as the image elastic charge fractionalizes in the presence of sharp corners, with partial disclinations localizing at the inner corners of the hole. To test this idea further we repeated the simulation for the buckling of triangular and circular frames, subjected to the forces shown in \figref{fig:Triangle}. We find that in the triangular case the induced elastic charges on the boundary are only partly fractionalized. One positive disclination monopole is localized near the corner, whereas the other is smeared on the opposite side of the triangle (\figref{fig:Triangle}(a) bottom and top sides). The negative disclination monopoles are also localized and attracted to the two corners of the triangle's base. In the circular case (\figref{fig:Triangle}(b)) we find that the induced elastic charges do not fractionalize - they are smeared evenly over the boundary instead.

\begin{figure}
	\includegraphics[width=0.95\linewidth]{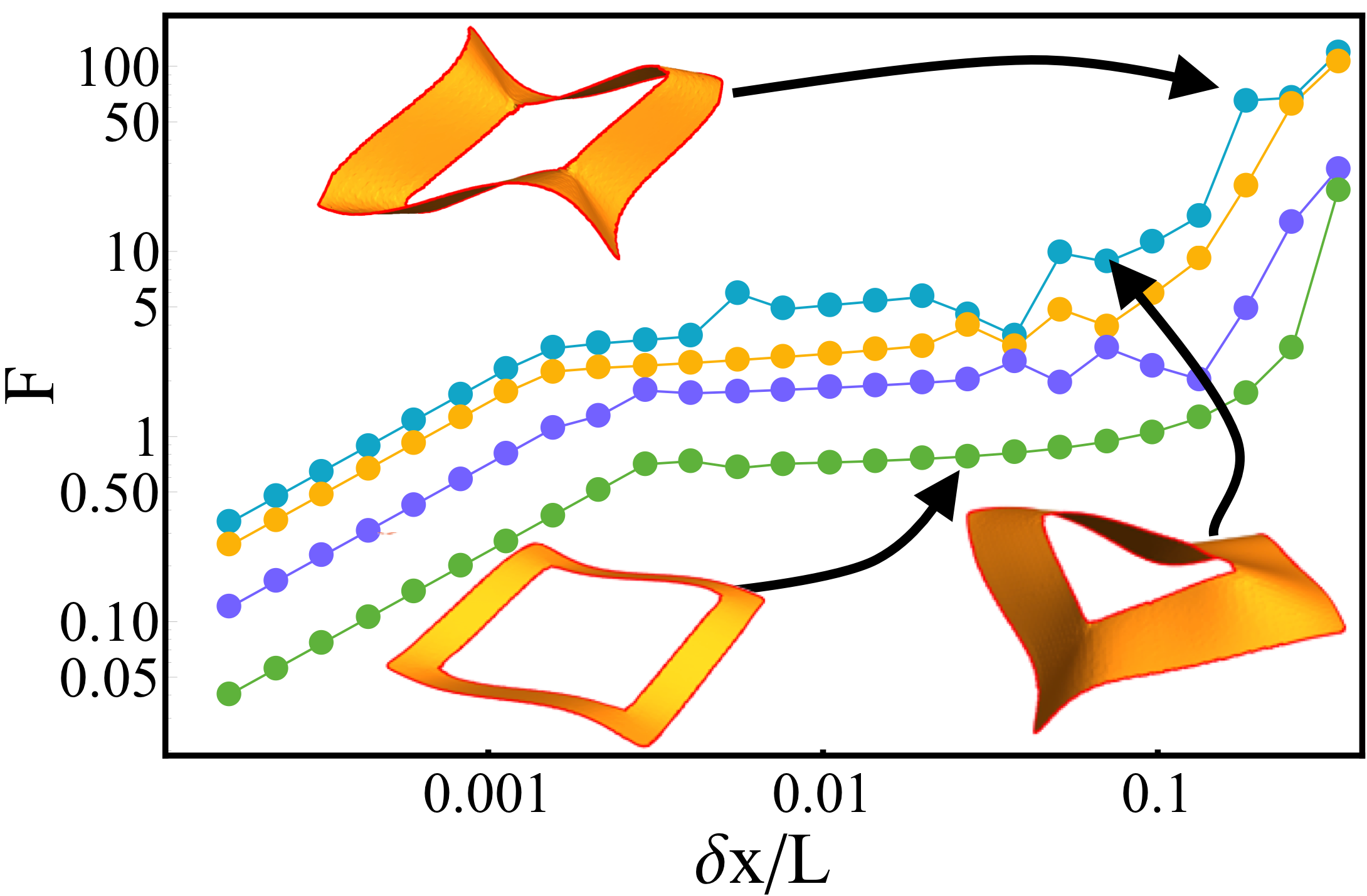}
	\caption{Force-displacement curves plotted on a log-log scale for frames with $L=1$, $t=0.005$. The frame width ranges from top to bottom across $w=0.25,0.225,0.175,0.125$. Different modes of buckling (see Fig.~\ref{fig:Illustration}) are responsible for slight qualitative differences in the buckling pathways, depending on the ratio $w/L$.}
	\label{fig:ForceDispConf}
\end{figure}

In \figref{fig:ForceDispConf} we plot force-displacement curves on a double-log scale for frames of fixed thickness and four different frame widths. The response is linear for small displacements, suggesting that the frame remains planar and responds by stretching controlled by the Young's modulus $Y$. A ``plateau'' with much smaller slope then develops,  corresponding to buckling. In this regime the Young's modulus is effectively replaced by  $\kappa/L^2$, as discussed earlier. This is a dramatic softening for easily bendable frames. Finally, when the displacement is comparable to the hole size, the response stiffens and stretching becomes important once again.
The ``noisy'' results (especially in the top curve of \figref{fig:ForceDispConf}), are not numerical errors, but rather correspond to locally stable configurations as shown in the inset. The small energetic differences between these different buckling modes reflects weak interactions between the screening partial disclinations.
Although the energetic differences are small, we find that for frame sizes $w/L \leq 0.175$ the $+:L/L:-$ configuration (\figref{fig:Illustration}(d)) is favorable, whereas for wider frames the $+:L/L:+$ configuration is favorable (\figref{fig:Illustration}(f)).

\begin{figure}
	\includegraphics[width=1\linewidth]{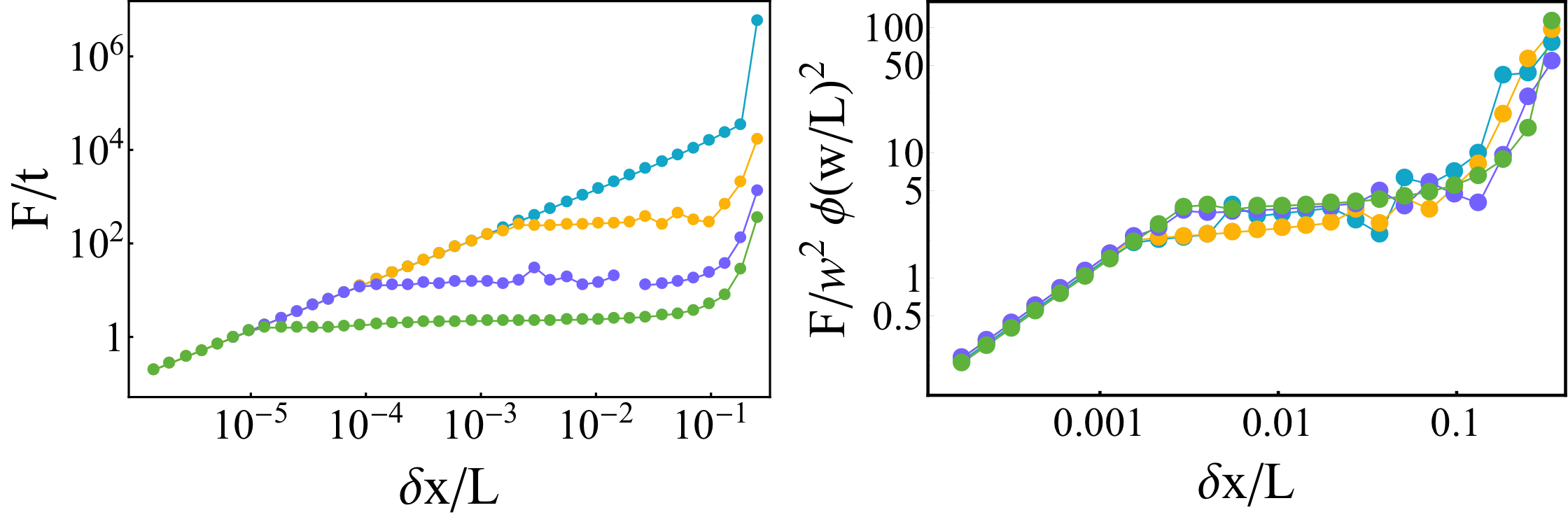}
	\caption{Log-log force-displacement plots: (a) fixed frame size $w/L=0.35$ and thicknesses $t = 0.0005$ (green), $0.0011$ (blue), $0.005$ (orange) and $0.05$ (cyan),  with the force normalized by the thickness (b) fixed thickness $t = 0.005$ and four frame widths $w/L = 0.5,0.45,0.35,0.25$, with the force normalized by the $w/L$-dependence according to \eqref{eq:keffBuckledSmall}.}
	\label{fig:ForceDispNorm}
\end{figure}

Force-displacement curves for two different protocols are shown in \figref{fig:ForceDispNorm}. In \figref{fig:ForceDispNorm}(a) four different thicknesses (or equivalently four different FvK numbers) are shown at fixed frame width. The force is normalized by the thickness and the data collapses to a single linear curve until the onset of buckling. Frames with a larger thickness buckle at larger values of the displacement as expected. \figref{fig:ForceDispNorm}(b) describes frames with fixed thickness and variable frame width. Using \eqref{eq:KeffPlanar}, the force is normalized by $w^2 \Phi\brk{w/L}^2$, and collapses to a single curve until the onset of buckling. The nontrivial geometric dependence of the force response is well captured by a single image quadrupolar charge, embodied in the function $\Phi(w/L)$ described by \eqref{eq:InducedDisclination}; the post-buckling collapse in \figref{fig:ForceDispNorm}(b) strongly suggests that image charges provide an accurate description even for buckled frames.
\begin{figure}
	\includegraphics[width=1\linewidth]{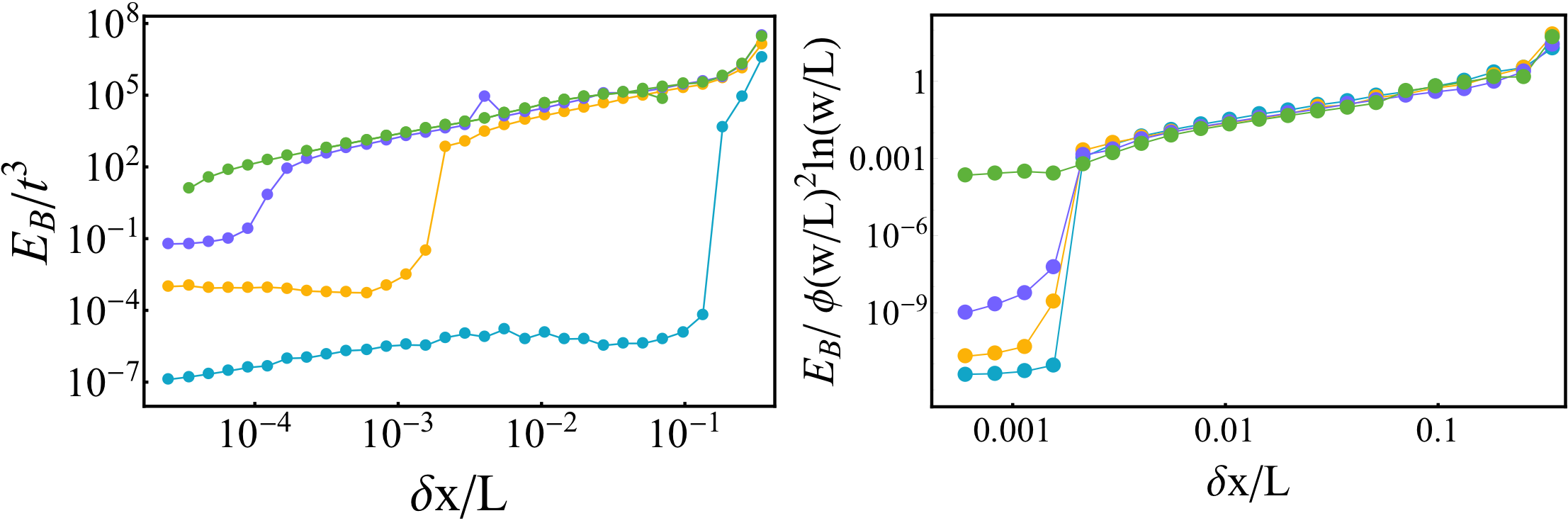}
	\caption{Bending energy as a function of displacement for (a) fixed hole size and different thicknesses, normalized by $t^3$ (which scales with the bending rigidity $\kappa$), and (b) fixed thickness and different hole sizes as in \figref{fig:ForceDispNorm}(b), normalized according to \eqref{eq:keffBuckledSmall}. The parameters here are identical to those used in \figref{fig:ForceDispNorm}.}
	\label{fig:BendEnNorm}
\end{figure}

The departure of the force response curves from linearity, as in \figref{fig:ForceDispNorm}, is a clear but indirect signature of buckling out of plane. While in experiments this is  a useful approach for estimating the buckling transition, numerical simulations provide easy access to the bending energy of frames, which is a direct measure of the amount of buckling.
\figref{fig:BendEnNorm}(a) shows a plot of the bending energy normalized by $t^3$ ($\kappa\sim t^3$), for a fixed frame size and different thicknesses. The collapse to a single curve in the post-buckled regime clearly shows the dominance of bending there and reflects the significant screening of the image charge by Gaussian curvature. The jump in each curve clearly identifies the buckling transition.
In \figref{fig:BendEnNorm}(b), the thickness is kept fixed and the frame size is varied. Now using \eqref{eq:keffBuckledSmall}, we normalize the bending energy by $\Phi(w/L)^2 \ln(w/L)$, which shows a collapse onto a single curve in the post-buckling region. 

\begin{figure}
\includegraphics[width=\columnwidth]{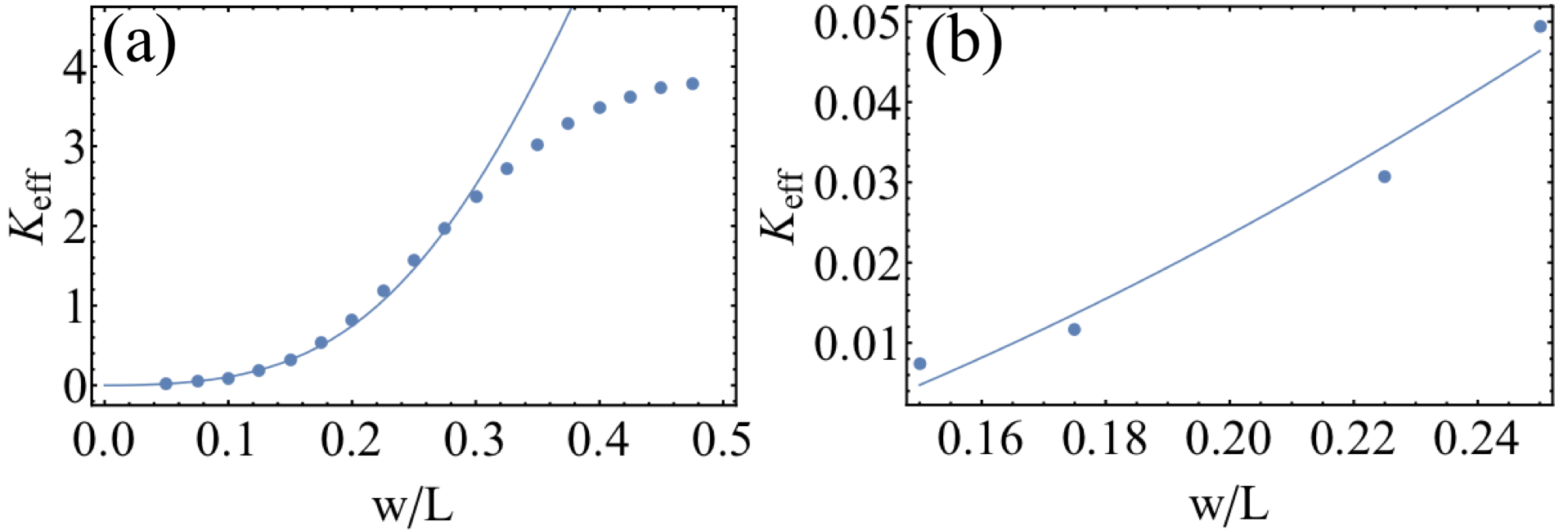}
\caption{Effective spring constant $k_\text{eff}=\dif f/\dif\delta x$ as function of the dimensionless frame width $w$ for  (a) planar and (b) buckled frames with $L=1$ and $t=5 \times 10^{-4}$. The blue dots are from the numerical simulations. The solid lines are analytical results corresponding to \eqref{eq:KeffPlanar} in (a) with no fitting parameters, and to \eqref{eq:keffBuckledSmall} in (b) with an overall prefactor as a fitting parameter.}
    \label{fig:KeffJoined}
\end{figure}

The analytical approach in the planar problem provided us with an explicit expression for the effective spring constant for all values of $w/L=(1-H/L)/2$. The assumption of energy localization in the vicinity of the inner corners (as appropriate for a square frame), was qualitatively confirmed in \figref{fig:PlanarConf} and allows us to estimate the energy of the $3d$ configuration.
To quantitatively test this physical picture, we compare in \figref{fig:KeffJoined}(a) the effective spring constant extracted from the set of planar frame simulations with the analytical result in \eqref{eq:KeffPlanar} estimated by four planar partial disclinations at each corner. With no fitting parameters we find excellent agreement between the two, for frames with $w/L<1/4$ (or equivalently $H/L>1/2$), as expected. The exact expression for $k_{\mathrm{eff}}=\dif f/\dif\delta x$ for a planar frame and \emph{arbitrary} frame width is given in \appref{app:keff}.

In \figref{fig:KeffJoined}(b) we plot the effective spring constant, extracted from the simulations in the buckled regime, as a function of the normalized frame width $w/L$, together with the analytic calculation in \eqref{eq:keffBuckledSmall}. We find good agreement with only one overall fitting prefactor. This prefactor reflects the difference between the (non-topological) screening disclination and the topological disclination studied for example in Ref.~\cite{Seung88}.

\begin{figure}
\includegraphics[width=\linewidth]{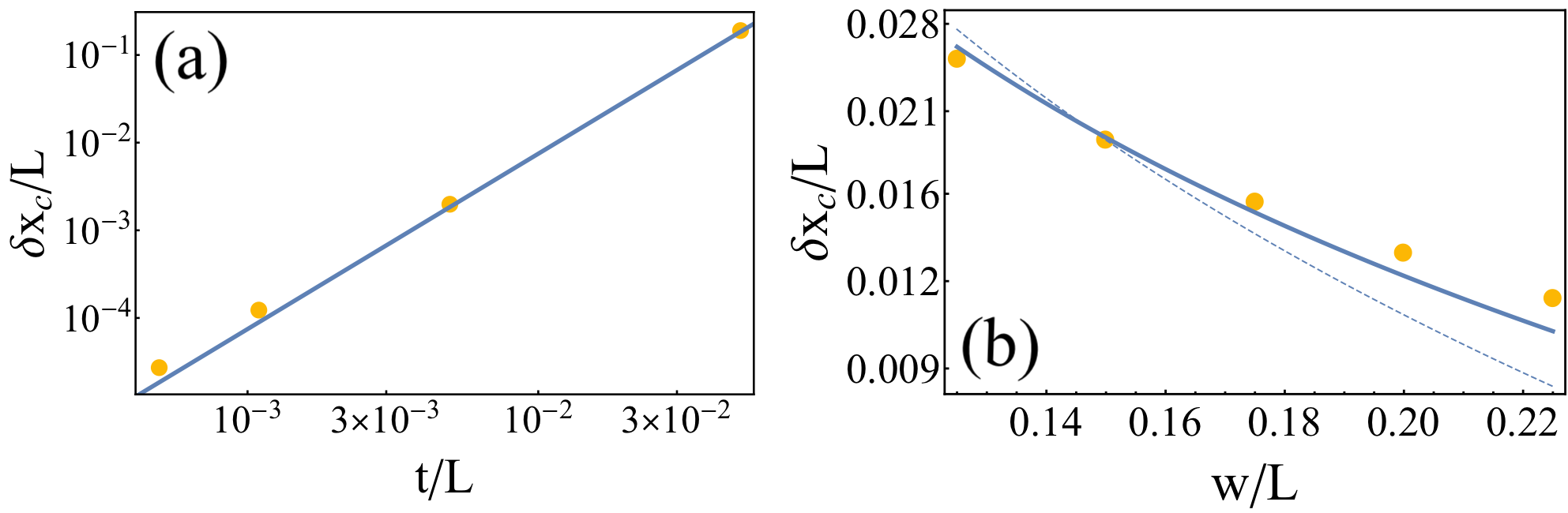}
\caption{Critical displacement for buckling as function of (a) the thickness and (b) the frame width, with solid lines fitted to \eqref{eq:CritDisp}. The dashed line in (b) goes as $w^{-2}$, corresponding to the na{\"i}ve approximation $\Phi(w/L)=\mathrm{constant.}$}
	\label{fig:deltaXcrit}
\end{figure}

Finally in \figref{fig:deltaXcrit}(a) we plot the critical displacement for buckling as a function of the thickness on a log-log scale. The straight line $\propto t^2$ superimposed on the numerical data confirms the scaling prediction given in \eqref{eq:CritDisp}, with $\delta x_c/L \approx 75 t^2$.  From \eqref{eq:CritDisp} we then find that  $\gamma^{\mathrm{frame}}_c /12 w^2 \Phi(w/L) = 75$, which, along with $w/L=0.35$, gives $\gamma^{\mathrm{frame}}_c \lesssim 80$. This is smaller than the value of $\gamma_c$ obtained for the buckling of topological disclinations ($\gamma_c \approx 100 - 120$). This difference between the two highlights the distinction between the buckling of non-topological partial disclinations and of topological ones. A detailed investigation of such non-topological partial disclination is left for future work.
The critical displacement extracted from simulations as a function of $w/L$ for a fixed thickness $t/L = 0.05$ is plotted in \figref{fig:deltaXcrit}(b). Within the ``na{\"i}ve'' approach, where the screening disclinations are assumed to be independent of the frame width {(i.e. $\Phi(w/L)=\mathrm{const.}$)}, the critical displacement is expected to scale like $1/w^2$, as shown by the blue dashed curve. Taking into account the function $\Phi(w/L)$, as in \eqref{eq:CritDisp}, we obtain the solid blue line which agrees well with our numerical results.

\section{Summary and discussion}
\label{sec:summary}
In this paper we have applied the geometric formulation of elasticity to study the mechanics of square frames, especially their buckling into the third dimension. Using the notion of external force or displacement-dependent elastic charges, we developed tools for an explicit and direct computation of the stressed state of planar frames given an arbitrary hole geometry.
We have shown how stress-induced image charges within the hole can fractionalize into partial disclinations localized at sharp corners of the hole. In the buckled regime, frames respond to applied loads by screening these induced image charges. The bending energy of elastically distorted frames was then estimated within this formalism. The image charge approach made the challenging nonlinear problem of post-buckling mechanics accessible by directly relating it to a simpler pre-buckling computation within the planar problem.
%Using this tool we could compute the induced image elastic charge formed within the hole, and use this result to estimate the bending energy of the buckled frame as composed of non-interacting partial disclinations screening the image charge form in the planar problem.

The specific simplifications afforded to us by the square geometry of the hole allowed analytical predictions to be quantitatively tested against numerical simulations. Our findings support the analysis both qualitatively and quantitatively. The analogy with electrostatics provides an appealing intuitive picture which allows for the interpretation of the mechanics of frames through the formation of elastic charges.
%and their distribution over the frame's edge, very similar to electrostatic charges, localizes at the vicinity of sharp corners.

It seems plausible that the mechanics of more elaborate kirigami structures under stress could profitably be thought of as a problem of interacting elastic charges. This perspective could provide a powerful organizational framework to think about the mechanics of kirigami meta-materials. In this regard, our work can be seen as the first step towards addressing the general problem. We leave a detailed study of kirigami with many interacting charges (either screening or fictitious) for future investigation.
%, the case of coupled frames, can be described as a problem of interacting charges, wither imaginary or screening. In fact, the Kirigami pattern shown in \figref{fig:Illustration} support this observation. When stretched along the $x$ direction, the screening disclinations form closely packed quadrupolar structures on the junctions between the holes. This results with an attractive interaction energy, which therefore support buckling. 
%A rigorous mapping of the problem of Kirigami into that of interacting charges is left for a future investigation.

\section{Acknowledgments}
We thank Itai Cohen and Paul McEuen for insightful discussions.
Work by M.J.B. was supported by the KITP grant PHY-1125915 and by the NSF DMREF program, via grant DMREF-1435794. Work by D.R.N. was primarily supported through the NSF DMREF program, via grant DMREF-1435999, as well as in part through the  Harvard Materials Research and Engineering Center, via NSF grant DMR-1420570. M.M. acknowledges the USIEF Fulbright program. M.M., S.S, and M.J.B. thank the Syracuse Soft \& Living Matter Program for support and the KITP for hospitality during completion of some of this work.

\appendix
\section{Image charge method for a sheared planar annulus}
\label{app:annulus}
Denote by $\mathcal{C}\brk{r}$ an undeformed circular domain of radius $r$. We define the annular domain with outer and inner radii $\rout$ and $\rin$  by $\Omega = \mathcal{C}\brk{\rout} \backslash \mathcal{C}\brk{\rin}$. The width of the annulus is denoted by $\delta r = \rout-\rin$.The plane-stress problem in an annular geometry consists of solving the bi-harmonic equation $\Delta \Delta \chi = 0$ (see  \eqref{eq:biharmonic}) in $\Omega$, with boundary conditions
\begin{equation}
\begin{split}
&\sigma^{xy}|_{\rout} = \sigma^{yx}|_{\rout}  = \sigma_0\\
&\sigma^{xx}|_{\rout} = \sigma^{yy}|_{\rout}  = 0 \\
&\sigma^{rr}|_{\rin}  = \sigma^{\theta\theta}|_{\rin}= 0\\
&\sigma^{r\theta}|_{\rin}  = \sigma^{\theta,r}|_{\rin}= 0,
\end{split}
\end{equation}
on the outer and inner edges respectively.

\begin{figure}
	\centering
	\includegraphics[width=0.5\linewidth]{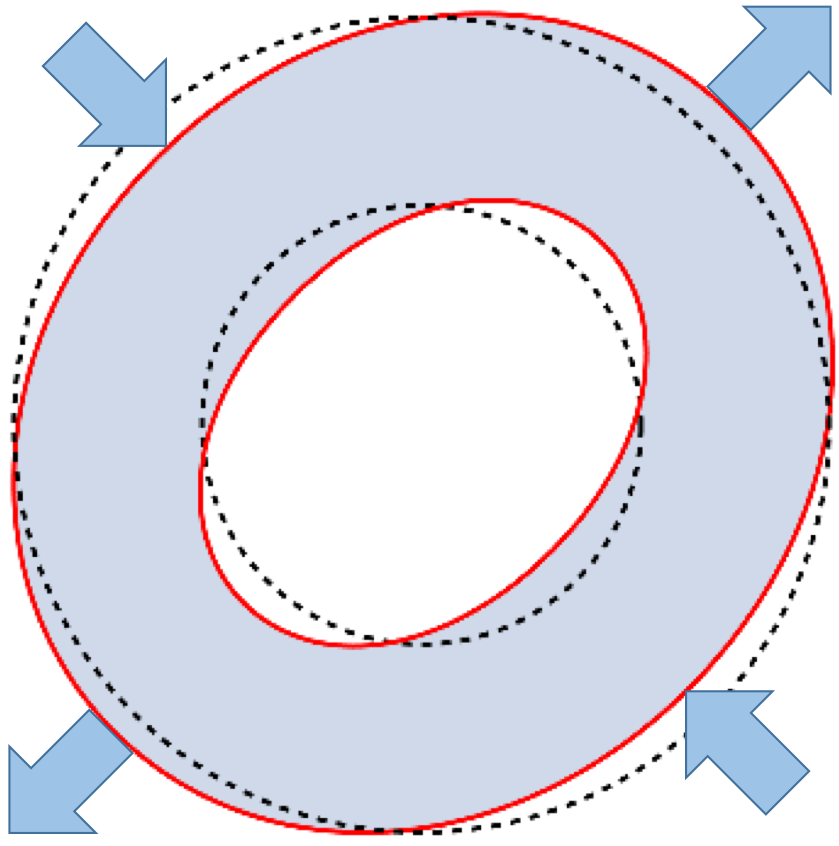}
	\caption{The analytic solution for an annulus subjected to pure external shear on its external boundary.}
	\label{fig:annulus}
\end{figure}

The solution for the stress function is given by \cite{Barber1992}
\begin{equation}
\begin{split}
\chi\brk{r,\theta}&= \brk{a r^4 + b r^2 + c +d r^{-2}} \sin 2\theta,
\end{split}
\label{eq:SFsolution}
\end{equation}
with
\begin{equation}
\begin{split}
a &= \frac{2 \rout^2-3 \rin^2}{12\brk{\rout^2-\rin^2}^2} \sigma_0 \\
b &= \frac{2\rin^2 - \rout^2}{4\brk{\rout^2-\rin^2}^2} \rin^2 \sigma_0\\
c &= -\frac{\rin^6}{4\brk{\rout^2-\rin^2}^2} \sigma_0\\
d &= \frac{\rin^6 \rout^2}{12\brk{\rout^2-\rin^2}^2} \sigma_0.
\end{split}
\label{eq:Coeff}
\end{equation}
The equilibrium configuration is plotted in \figref{fig:annulus} .
Instead of solving \eqref{eq:biharmonic} in $\Omega$, the same solution can be obtained by solving
\begin{equation}
\frac{1}{Y}\Delta \Delta \chi = \Kim,
\label{eq:biNonharmonic}
\end{equation}
in the disk $\mathcal{C}(\rout)$, where $\Kim$ can be non-zero in $\mathcal{C}(\rin)$, and corresponds to the image elastic charges required to satisfy the boundary conditions.

The first two terms ($a$ and $b$) in \eqref{eq:SFsolution} can be viewed as resulting from image charges at infinity. The remaining two terms ($c$ and $d$) can be viewed as being induced by a singular source term at the origin of the form
\[
\Kim = 2Q_\text{Im}\partial_x\partial_y\delta\brk{\xvec} + 2H_\text{Im}\partial_x\partial_y\Delta\delta\brk{\xvec}
\]
with the magnitude of the charges given by 
\begin{equation}
\begin{split}
Q_\text{Im} =  -\frac{ \pi \rin^6 \sigma_0}{Y \brk{\rout^2-\rin^2}^2} , \\  H_\text{Im} = -\frac{ \pi \rin^6 \rout^2 \sigma_0}{6 Y \brk{\rout^2-\rin^2}^2} . 
\end{split}
\label{eq:ImQuad}
\end{equation}
The two singularities correspond to a fictitious quadrupolar charge and a fictitious hexadecapolar charge, respectively.

\begin{figure}
	\includegraphics[width=0.75\columnwidth]{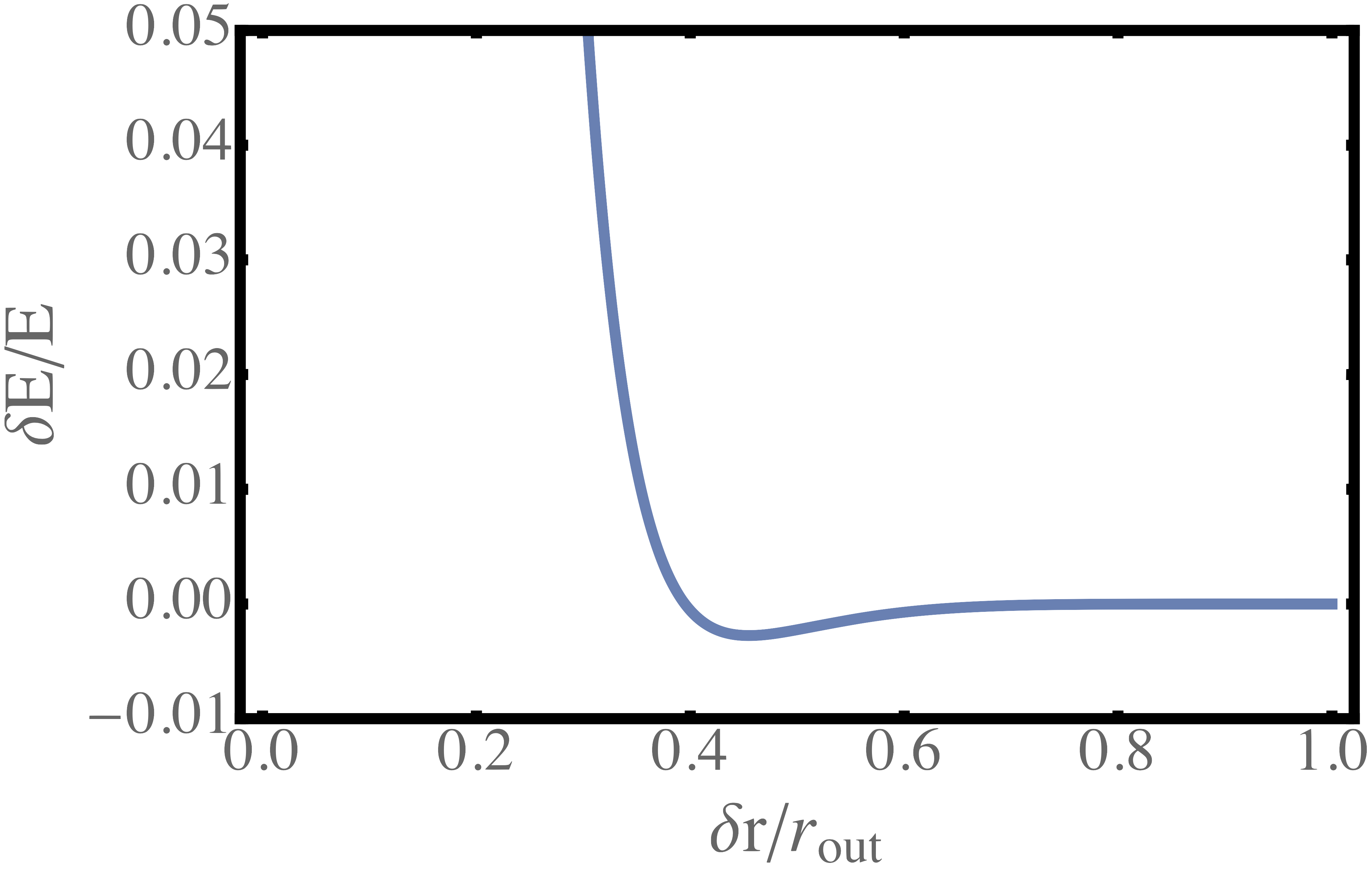}
	\caption{The deviation from the exact solution of the approximate energy of a sheared circular frame, obtained by neglecting the hexadecapole. The relative deviation is less than $5\%$ up to a relative hole size of $\eta = \rin/\rout = 0.7$.}
	\label{fig:NoOctapole}
\end{figure}

To determine the range of validity of the quadrupole approximation, for the case of a prescribed shear force $\sigma_0$, we calculate the energy $E_\text{annulus}$ of a sheared annulus from \eqref{eq:SFsolution}. The energy is composed of the  boundary terms, the quadrupolar and hexadecapolar terms, and an interaction term. Summing them gives
\begin{equation}
E_\text{annulus} =  \frac{12 - 12 \eta^2 + 3 \eta^4 + 5 \eta^6}{24 \brk{1 - \eta^2}} \frac{\sigma_0^2}{Y}\pi \rout^2 ,
\label{eq:CircEn}
\end{equation}
where $\eta=\rin/\rout = 1-(\delta r/\rout)$ is the relative hole size. In the limit $\delta r \to \rout$, which equivalent to $\eta \to 0$, the domain is a full disk, and this expression recovers the standard result $E_\text{disk} = \frac{\pi \rout^2}{2 Y} \sigma_0^2$.
We now compare the exact solution of the circular geometry with an approximation obtained by neglecting the hexadecapolar term. In \figref{fig:NoOctapole} we plot the relative deviation of the approximate solution from the exact one, as a function of the hole size. We find that the deviation is less then $5\%$ for frame widths of size $0.3 < \delta r/\rout$, which provides a guide for the pure quadrupole approximation applied to circular frames. Indeed, upon referring to \figref{fig:Illustration}(c), we expect the quadrupole approximation to be adequate when $0.3 \lesssim w/L $.

\section{Curvature distribution over the boundaries}
\label{app:curvarture}

There are two different approaches to understanding the notion of elastic charge distributions on boundaries. We distinguish between elastic charges in the planar and buckled $3d$ problems.

\subsection{Elastic charges - the planar problem}
\begin{figure}
	\centering
	\includegraphics[width=0.7\linewidth]{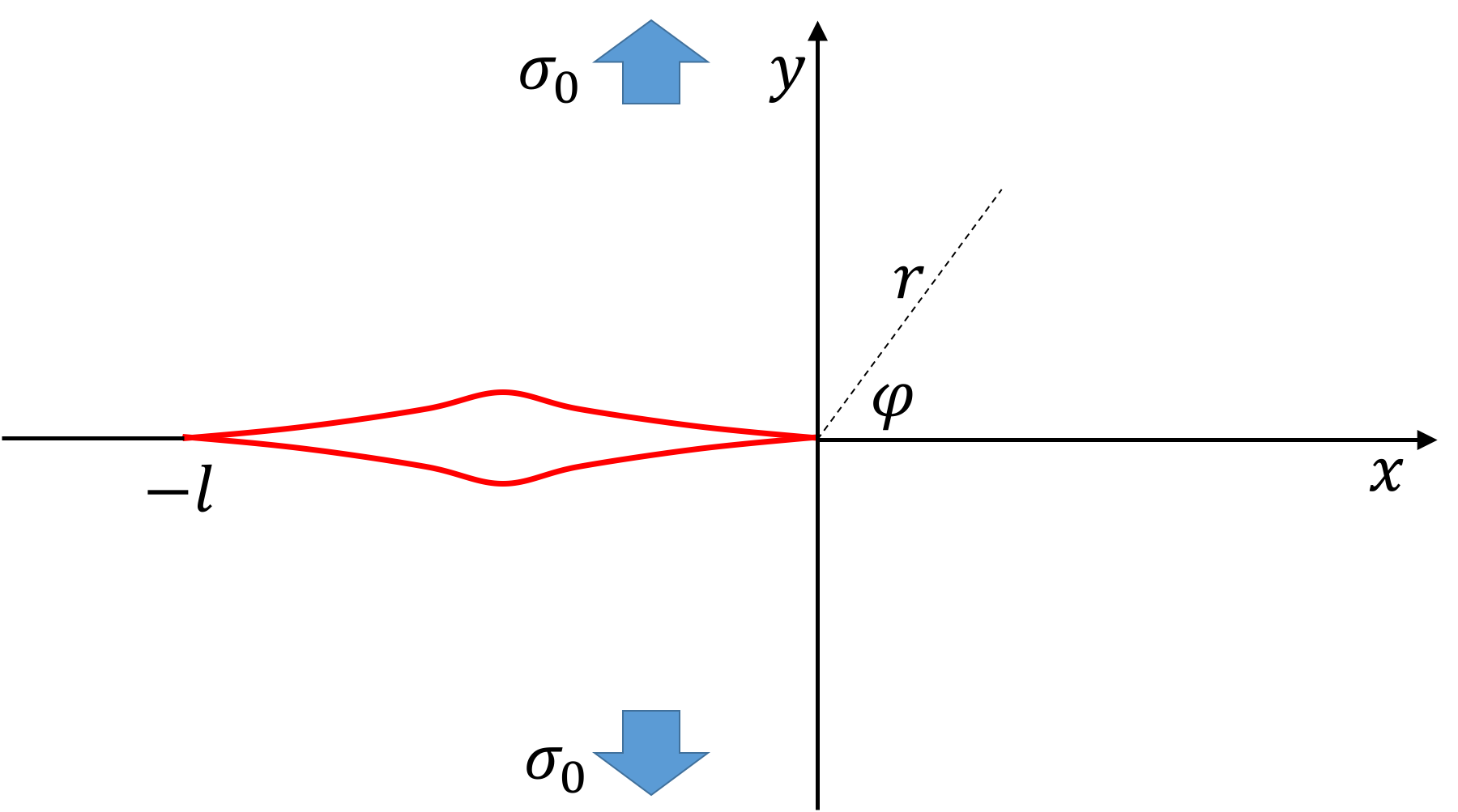}
	\caption{A finite slit of length $l$ in an infinite medium subjected to a remote stress $\sigma_0$. When loaded, the slit lips are opened, and the stress field becomes singular near the slit ends.}
	\label{fig:crackillustration}
\end{figure}

For planar problems one can understand the basic idea using the equation for the stress function. In this case $\chi$ satisfies,
 \[
\frac{1}{Y} \Delta \Delta \chi  = -\Kim
 \]
 where $\Kim$, the charge induced by the external load, is expected to be realized by the elastic charge distribution over the boundary. We set $\phi \equiv -\Delta \chi$ and recover Poisson's equation $\Delta \phi  = Y \Kim $. Gauss' law, with $\phi$ treated as an electric potential, determines the charge on the boundary. 
 As an example, consider the problem of a finite slit located on the $x$ axis at $-l<x<0$ (see \figref{fig:crackillustration}). 
 When uniaxial remote loads are prescribed along the $y$ direction, the solution to the bi-harmonic stress function that satisfies stress free boundary conditions along the slit, near the right tip, is
\begin{equation}
\chi = A \, \sigma_0 \sqrt{l} \, r^{3/2} \brk{\cos \frac{\varphi}{2}  + \frac{1}{3}\cos \frac{3\varphi}{2}} + O(r^{5/2}),
\label{eq:CrackSol}
\end{equation}
with $A$ a positive numerical constant, and $\sigma_0$ the remote stress applied along the $y$ direction. Here $(r,\varphi)$ are polar coordinates measured from the slit tip, with $\varphi = 0$ pointing in the $x$ direction.
In Cartesian coordinates $\phi \equiv -\Delta \chi = -(\sigma^{xx}+\sigma^{yy})$, which means that $-\phi$ is the pressure.  
Close to the slit tip (see \cite{Sun2011}) \eqref{eq:CrackSol}   yields 
\begin{equation}
\phi \approx \frac{\sqrt{l}\, \sigma_0}{\sqrt{2 r}} \cos \frac{1}{2} \varphi
\end{equation}
Calculating the normal derivative along the slit, Gauss' law with the ``potential'' $\phi$, gives a charge density 
\[
\rho_\text{Im} \brk{r} =- \frac{A \, \sqrt{l}\, \sigma^0}{Y r^{1/2}} + O(r^{1/2}).
\]
This result shows that there is a negative power-law singularity near the corner. This is consistent with our image charges approach and with the singularity of electric charge density at the tip of a 2d conductive needle.
We expect similar power-law singularities near the frame corners for other hole shapes.

\subsection{Curvature charges - the 3d problem}
\begin{figure}
	\centering
	\includegraphics[width=0.8\linewidth]{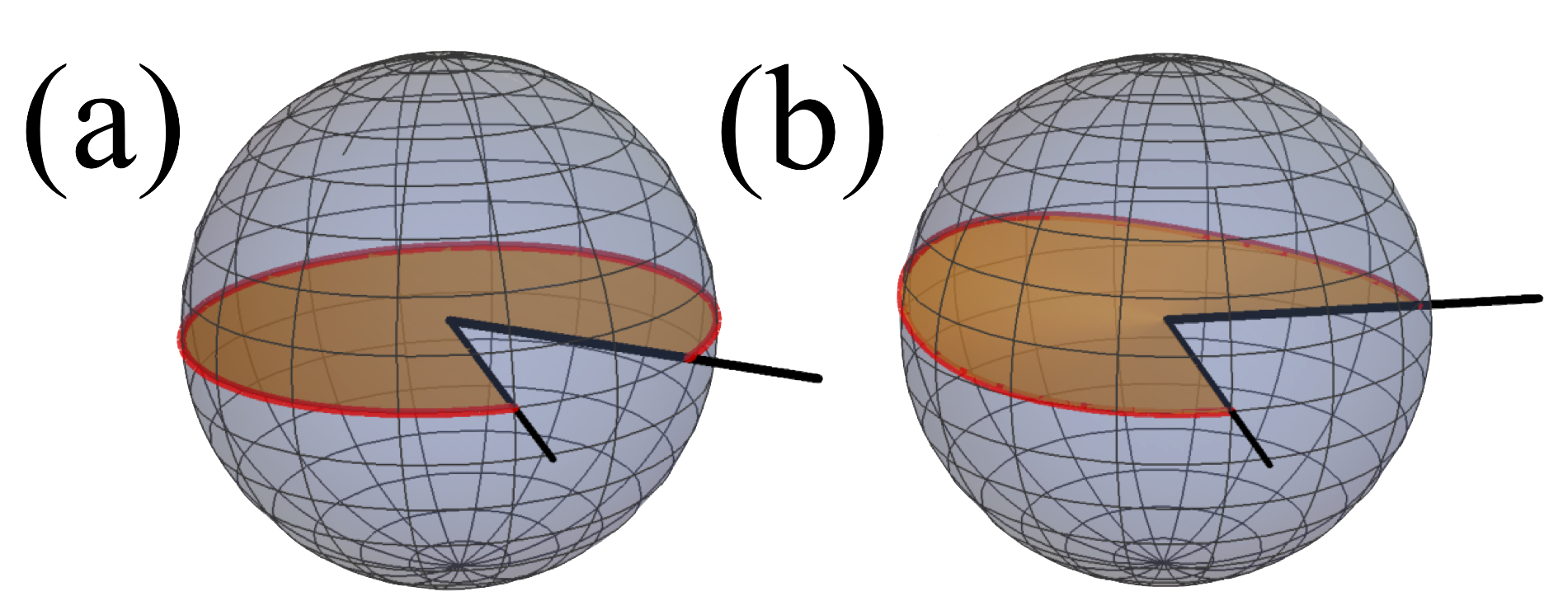}
	\caption{Illustration of buckled partial disclinations prescribed by fixing an angle between two rigid rods attached to the sector's radial edges. (a) Unstressed sector of angle deficit $\Delta = \pi/3$; (b) a negative buckled conical configuration induced by opening the rods by an angle $\delta \theta$}
	\label{fig:pds}
\end{figure}

In the 3d problem there are now curvature charges in addition to the elastic charges introduced in the planar case. These correspond to singular Gaussian curvature of the configuration. The equation for the stress function is then
\[
\frac{1}{Y} \Delta \Delta \chi  = K - \Kim.
\]
Since $K$ and $\Kim$ can, in principle, cancel each other, we expect $K$ to distribute over the boundary to cancel $\Kim$ in the inextensible limit (very thin sheet).
Although the Gaussian curvature is well-defined in the bulk, its meaning on the boundary is not immediately clear. To clarify this point we focus on the deformation illustrated in \figref{fig:pds}. In \figref{fig:pds}(a), the sector is flat, and the tangent vector to the boundary is everywhere continuous, except of the two intersection points between the circular edge with the radial edges, and the point of intersection between the two radial edges. These discontinuities contribute to the geodesic curvature, and  flatness ($K=0$) together with the Gauss-Bonnet theorem:
\begin{equation}
\int_{\Omega} K \text{d}  S  + \oint_{\partial \Omega} k_g \text{d}  l = 2 \pi.
\label{eq:Bonnet1}
\end{equation}
lead to
\begin{equation}
\oint_{\partial \Omega} k_g \text{d}  l = 2 \pi . 
\label{eq:Bonnet2}
\end{equation}
Upon isometrically deforming the sheet as illustrated in \figref{fig:pds}(b), i.e. by introducing slight opening angle between the radial edges, the geodesic curvature is everywhere conserved, except at the corner between the radial edges.
Assuming the angle excess between the radial edges is $\Delta \theta$, one finds
\[
\oint_{\partial \Omega} k_g \text{d}  l = 2\pi + \Delta \theta.
\]
The Gauss-Bonnet theorem then requires that $K$ be modified by the imposed deformation. The deformation being isometric also forces $K$ to be singular, vanishing everywhere but at the origin,
\begin{equation}
\int_{\Omega} K \text{d}  S = -\Delta \theta,
\label{eq:Bonnet3}
\end{equation}
which gives
\[
K =  -\brk{\Delta\theta} \delta^{(2)}\brk{\xvec}.
\]
This argument shows that a singular source of Gaussian curvature can be related to the excess or deficit geodesic curvature along the boundary.

\section{Calculation of $k_\text{eff}$}
\label{app:keff}
In the body of the paper we transformed the problem of a planar frame under load,as expressed in \eqref{eq:FullProblem}, to a simpler problem as expressed in \eqref{eq:EnDiscrete} and \eqref{eq:lambda}. The calculation of the unknown charges requires an integration over the frame of the terms in \eqref{eq:EnDiscrete}.
The functions $\phi_1$, $\phi_2$ and $\Phi$ are expressed in \eqref{eq:phi1}, \eqref{eq:Qdeltax}, and \eqref{eq:InducedDisclination} in terms of the energy minimizing charges.
The effective spring constants may then be expressed in terms of $\Phi$.
The calculation for the unknown charges, the functions $\phi_1$, $\phi_2$, $\Phi$, and the effective spring constants, are all presented in detail in the attached \textit{Mathematica} notebook.

\bibliographystyle{apsrev4-1}
\bibliography{references}

\end{document}